\documentclass[prd,superscriptaddress,amsfonts,amssymb,amsmath]{revtex4}
\usepackage{bm}
\usepackage{amsfonts}
\usepackage{latexsym}
\usepackage{graphicx}
\usepackage{amsmath}
\usepackage{palatino}
\usepackage{mathpazo}
\usepackage{textcomp}
\usepackage{float}
\usepackage{booktabs}
\usepackage{dcolumn}
\usepackage{booktabs}
\usepackage{multirow}
\usepackage{hyperref}
\hypersetup{colorlinks,citecolor=blue}
\usepackage{amsmath}
\usepackage{xcolor}
\usepackage{subcaption}
\usepackage{commath}
\def\jnl@style{\it}
\def\aaref@jnl#1{{\jnl@style#1}}
\def\aaref@jnl#1{{\jnl@style#1}}

\allowdisplaybreaks[1]

 \allowdisplaybreaks
\begin{document}
\title{Fuzzy Dark Matter Less-complex Wormhole Structures in Higher-Order Curvature Gravity}

\author{Z. Yousaf}
\email[Email: ]{zeeshan.math@pu.edu.pk}
\affiliation{Institute of Mathematics, University of the Punjab, Lahore-54590, Pakistan.}

\author{Kazuharu Bamba}
\email[Email: ]{bamba@sss.fukushima-u.ac.jp \textcolor{black}{(corresponding author)}}
\affiliation{Faculty of Symbiotic Systems Science,
Fukushima University, Fukushima 960-1296, Japan.}

\author{B. Almutairi}
\email[Email: ]{balmutairi@ksu.edu.sa}
\affiliation{Department of Mathematics, College of Science, King Saud University, P.O. Box 2455 Riyadh 11451, Saudi Arabia.}

\author{M. Z. Bhatti}
\email[Email: ]{mzaeem.math@pu.edu.pk}
\affiliation{Institute of Mathematics, University of the Punjab, Lahore-54590, Pakistan.}

\author{M. Rizwan}
\email[Email: ]{mrizwan.math@gmail.com}
\affiliation{Institute of Mathematics, University of the Punjab, Lahore-54590, Pakistan.}


\begin{abstract}
Fuzzy dark matter wormhole solutions coupled with anisotropic matter distribution are explored in higher-order curvature gravity. We derive the shape function for fuzzy wormholes and explore their possible stability. We study the embedding diagrams of the active gravitational mass associated with fuzzy dark matter wormholes by taking a certain shape function. Aiming to highlight the role of higher-order curvature gravity in the modeling of less complex fuzzy wormhole structures, we evaluate the complexity factor, the conservation equation, and null energy conditions. Our study reinforces more importance of uniformly distributed pressure effects throughout the less complex region than the emergence of energy density homogeneity in the stability of fuzzy wormholes. It is shown that the active gravitational mass of the fuzzy wormhole structures varies inversely with the radial distance, thereby suggesting the breaching of energy conditions at some arena of the Einasto index. Furthermore, it is revealed that stable fuzzy dark matter wormhole structures exist in nature in the surroundings of cold dark matter halos and galactic bulges. The important physics understood from our analysis is that in higher-order curvature gravity, feasible geometries of fuzzy dark matter wormholes exist naturally in the environments of different galactic haloes.
\end{abstract}
\maketitle

\section{Introduction}

There are very unique solutions to Einstein's field equations that illustrate the relationship between two extremely remote regions or parallel universes, among the many solutions that the equations have to describe cosmological events. Originally dubbed as the Einstein-Rosen bridge \cite{einstein1935particle}, these structures were first intended to use spacetime tunnels connected by electric lines of force to clarify fundamental particles like electrons, or more specifically, to answer the particle issue in general relativity (GR). However, the main issue was that these geometric constructions were not stable. This indicates that the bridge cannot stay open for an object (not even a photon) to pass through it. Following that, Morris and Thorne's \cite{morris1988wormholes} provided the necessary prerequisites for creating a traversable wormhole (WH) space-time (Wheeler \cite{wheeler1955geons} originated the term ``WH"). The WH concept is primarily topological and global, locally, it is described as a 2D spatial surface on an achronal hyper-surface referred to as a ``throat."
The Schwarzschild metric, which is maximally extended, is the most commonly encountered form of a WH metric. The intriguing and thrilling idea of time travel at least theoretically has been made possible by these things in a relatively short amount of time.

In the domain of GR, it is widely acknowledged that, for stable traversable WHs to exist, exotic matter must be present, which breaches the null energy condition (NEC). The logic of the implication of this condition comes from principles of topological censorship in the background of asymptotically flat regions, leading to the exploitation of constraints to some extent in the exploration of distant realms or alternate dimensions. It is well-known that a major challenge in the investigation of gravitational forces exists in the construction of traversable WH geometries using relativistic/exotic matter that acts per the energy criteria. Thus, it highlights the complex interaction between mathematical possibilities and feasible restrictions in the investigation of WHs. Various researchers have investigated and suggested extensions to the Einstein gravity models to avoid certain observational constraints of GR. An instance of this modification involves the consolidation of exotic matter as a scalar field that is nonminimally coupled in certain scalar-tensor theories of gravity. Furthermore, the gravitational amalgams in the theories that embody extra curvatures put forward alternative routes to cope with these challenges. This distinctive environment of gravitational theories describes our search to refine our perception of gravity and its implications \cite{ellis1973ether,bronnikov1973scalar,clement1981einstein,morris1988wormholes,visser1995lorentzian,hochberg1998dynamic}. Two distinct metrics are linked via a WH, making it impossible to conserve the flux in each manifold independently. This implies that events in one universe that are close to a WH will experience the effects of objects in the other spacetime. The combination of fuzzy dark matter (FDM) and higher-order curvature corrections provides a natural setting for WH geometries. FDM cores mimic Bose–Einstein condensates on kiloparsec scales (kpc) \cite{hui2017ultralight,miller2025fuzzy}, while theories like 4D EGB encode higher-order quantum corrections. FDM forms stable, compact solitons at galactic centers \cite{schive2014understanding}. These cores have densities $\sim 10^{9} M_{\odot}/\text{kpc}^{-3}$ and radii $\sim 100 \text{pc}$, creating extreme curvature conditions where WH throats could emerge. This synergy allows for the realization of traversable WHs without invoking artificially constructed exotic matter distributions. FDM’s quantum pressure (from its de Broglie wavelength) suppresses collapse to singularities, favoring smooth, horizonless geometries like WHs \cite{berezhiani2015reconciling}.

Di Grezia \textit{et al.} \cite{di2017spin} studied the traversable static WH structure in an Einstein-Cartan gravity. After analyzing energy conditions, they asserted that one can design traversable WHs with non-exotic matter. Dai \textit{et al.} \cite{dai2018new} explored a spherically symmetric WH structure intending to connect two finitely separated spatial points. They used two BH bodies at the antipodes in an environment of closed de Sitter metric demarcated by a matter shell. De Falco \textit{et al.} \cite{de2020general} established a criterion to differentiate a BH from WH after analyzing the corresponding equations for the spherically symmetric static spacetime. Dai and Stojkovic \cite{dai2019observing} proposed that the scalar, electromagnetic, and gravitational fields can be utilized to demonstrate this interesting phenomenon. De Falco \textit{et al.} \cite{de2021testing} studied various properties of WHs, including accretion disk and spectra to examine how modified gravity contributes to the presence of such structures. Dai \textit{et al.} \cite{dai2020form} proposed an easy way to understand WH formation due to the interaction of two massive stellar bodies. They asserted that relatively massive objects can be used to create WH geometry. The potential observational signature of WHs, and precision of astrophysical observations shortly \cite{simonetti2021sensitive,bambi2021astrophysical} will be our motivation to study such stellar structures. De Falco \textit{et al.} \cite{de2021reconstructing} found stable configurations of traversable WHs in a specific class of gravity theory. The mathematical strategy to understand WH existence in nature is also studied in \cite{de2021epicyclic,de2023epicyclic,ovgun2019exact,ahmed2025gravitational}.
De Falco \textit{et al.} \cite{de2023static} assumed a non-singular spherical WH solution and analyzed the flaring out and NEC, aiming to obtain viability constraints for the stable structure.

An isotropic pressure distribution in stellar structures implies a disparity between radial and tangential pressures. Factors like the phase transitions, solid core,  presence of multiple fluids, rotation, magnetic fields, viscosity, or superfluidity can introduce anisotropies in pressure within the fluid. Anisotropic solutions offer a plausible description of astrophysical entities, reflecting our comprehensive understanding of these objects. Lema\^{i}tre was the first to elucidate pressure anisotropy in a perfect fluid sphere, while Bower and Liang \cite{bowers1974anisotropic} highlighted the significance of locally anisotropic pressure equations (EOS) in relativistic contexts in 1974. In regions of extremely high density, such as  $\rho > 10^{16}$ kg/m$^3$, nuclear matter may exhibit anisotropic properties, as demonstrated in Ruderman's seminal work \cite{ruderman1972pulsars}. Several scholarly articles delve into the development of anisotropic solutions, underscoring the ongoing exploration of this phenomenon in astrophysical research \cite{astashenok2023chandrasekhar,malik2024study}. Recently,  In Ref. \cite{herrera2018new} Herrera explored self-gravitating compact anisotropic objects to analyze the notion of the vanishing complexity factor. In doing so, a general expression of the mass function was presented. By utilizing this mathematical approach, Herrera intended to describe the behavior of the locally anisotropic gravitational stellar objects and investigate the possible use of the vanishing complexity factor in this direction \cite{herrera2019complexity,herrera2020stability,contreras2022uncharged,herrera2024cracking,herrera2024irreversibility}.

Extended gravitational theories have been put forward to elucidate certain observable occurrences that General Relativity (GR) struggles to account for. Examples of such phenomena include cosmic accelerated expansion, dark energy, dark matter, massive pulsars, and super-Chandrasekhar white dwarfs, among others \cite{pretel2025white}. The gravity theories like Einstein-Gauss-Bonnet (EGB) gravity, $f(R)$ and $f(\mathcal{T})$ offer potential explanations for these phenomena (for more understanding, see Refs. \cite{riess1998observational,ade2015bicep2,eisenstein2005detection,jain2003cross,nojiri2011unified,sotiriou2010f,capozziello2011extended,capozziello2010beyond,joyce2015beyond,koyama2016cosmological,nojiri2017modified,yousaf2023quasi,de2010f,nojiri2021ghost, yousaf2025viscous}). The conventions that apply to matter generally do not hold in these situations. Instead, the geometry of the situation is taken
into account, effectively replacing the need for exotic matter. The efficient field theories and the terms involving higher-order
curvature is very important to accomplish this, which can be mathematically represented as extensions to the traditional Hilbert-Einstein
Lagrangian. By taking this approach, the properties of WHs can be mimicked, providing a potential solution to the problem at hand
\cite{ellis1973ether,bronnikov1973scalar,clement1981einstein,morris1988wormholes,visser1995lorentzian,hochberg1998dynamic}.

Black holes (BH) within $f(R)$ gravity models, exhibiting both constant and varying Ricci curvature, either in a vacuum or interacting with electrodynamics, have been documented in \cite{sebastiani2011static,multamaki2006spherically,hendi20142+,multamaki2007static,
nashed2020new,nashed2021analytic,nashed2019charged,de2009black,
jaryal2021gravitationally,eiroa2021thin,tang2021exact,yousaf2023generating}.
Meanwhile, scalar fields are coupled as the matter content in (3 + 1) and (2 + 1) dimensional $f(R)$ gravity scenarios in Refs. \cite{tang2021curvature,karakasis2021black,karakasis2021exact},
with further analysis of the resulting BH solutions. For cosmological applications, this specific $f(R)$ gravity framework in conjunction with nonminimally interacting scalar fields has already been investigated \cite{pi2018scalaron,de2016spotting}. The early studies of gravitational instability of scalar fields with self-interaction and structures, permit one to see insights into fuzzy dark matter BHs \cite{khlopov1985gravitational}.

The $f(R)$ theory led to the invention of WH solutions \cite{lobo2009wormhole}, in which the matter
within the WH obeys the energy criteria and the violation of these conditions is because of the higher-order curvature terms in $f(R)$ model of gravity. Bejarano \emph{et al.} \cite{bejarano2021thin}
studied thin shell WHs with spherical symmetry in (2 + 1) dimensional $f(R)$ gravity models,
where the Ricci scalar is constant. In addition, Ref. \cite{eiroa2016thin} presents an analysis of
the stability of thin-shell (3 + 1)-dimensional WHs in $f(R)$ model of gravity, including charge
and constant curvature. In the realm of quadratic $f(R)$ model, Lorentzian spherically symmetric
WHs with a constant scalar curvature have been discovered \cite{eiroa2016thin}. The Karmarkar
condition has been utilized to examine WH geometries in various $f(R)$ models \cite{shamir2020traversable}.
In Ref. \cite{chervon2021black}, WHs having kinetic curvature scalar are examined. Moreover, WHs in the massive
$f(R)$ gravity have been investigated using many redshift functions' behavior in \cite{tangphati2020traversable}.
In the field of gravitational physics, researchers frequently employ the concept of anisotropic pressure
as a fluid approximation when investigating self-gravitational fluid structures. Solving the gravitational
equations related to non-static and static anisotropic spheres of fluid has been the subject of many studies.
Many studies have been conducted on the hydrodynamical consequences of pressure showing local anisotropy and
its applicability in the context of altered gravitational models. Anisotropies in fluids can result from a
variety of things, including rotation, different kinds of phase transitions, magnetic fields, II-type superconductors,
viscosity, solid cores, and P-type superfluids
\cite{mustafa2021wormhole,astashenok2023chandrasekhar,malik2024study,naseer2024decoupled, naseer2024role, naseer2024charged, naseer2024estimating, naseer2024complexity}.

The study to explore WH geometries has been a central point in modified theoretical theories, like Einstein-Gauss-Bonnet (EGB) and Lovelock theories \cite{mehdizadeh2015einstein,kanti2012stable,maeda2008static}. It is noteworthy that the EGB theory has been a source of research by many researchers to understand various hidden features of our enigmatic universe \cite{nojiri2005modified,nojiri2024propagation,elizalde2023propagation,luo2024constraints}. By analyzing WHs in the EGB model, one may investigate how higher-dimensional curvature factors affect spacetime structure. This theory was initially put forward by Lanczos and later independently redeveloped by David Lovelock. Extensively researched, the EGB gravity is a subject of significant study due to its derivation in String theory's low-energy domain, its stability when perturbed around flat space-time, and its capacity to generate nontrivial gravitational self-interactions without ghost-like instabilities \cite{nojiri2019ghost,dehghani2009lorentzian,zangeneh2015traversable,lanczos1938remarkable,lovelock1972four}. When examining cosmic perturbations, it is known that the $F(R,\mathcal{G})$ model of gravity suffers from ghost degrees of freedom, which can appear at many levels of the theory, including as perturbative propagating modes. In Ref. \cite{nojiri2021ghost}, a ghost-free $F(R,\mathcal{G})$ model of gravity was examined.

Since the GB Lagrangian is a total derivative and does not affect gravitational dynamics, the EGB gravity displays topological features in 4D. To have non-trivial gravitational effects within the EGB theory, a dimensionality of $D \geq 5$ is necessary. Glavan and Lin \cite{glavan2020einstein} recently resolved this limitation by adjusting the GB parameter $\alpha$ to $\frac{\alpha}{D - 4}$, allowing for significant dynamics in four dimensions when approaching the limit $D \rightarrow 4$. To be specific, it is useful to consider it as the $4D$ EGB gravity \cite{kumar2020rotating,haghani2020growth,fernandes20224d,zanoletti2024cosmological,nashed2023isotropic}, notable for its unique feature of circumventing the implications of Ostrogradsky instability and Lovelock theorem. It is important to note upfront that Tomozawa explored dimensional regularization of this theory with comparable outcomes \cite{lovelock1971einstein,lanczos1938remarkable,zwiebach1985curvature}. The $4D$ EGB theory gained significant interest following the work of Glavan and Lin \cite{glavan2020einstein} (and referenced in), where they introduced static spherically symmetric vacuum black holes with unique characteristics. These black holes exhibit intriguing features such as singularity-free behavior, repulsive gravitational force at short distances, and the singularity is not reached by falling particles.

More than $80\%$ of the stuff in the cosmos is made up of DM, an unseen material. We are aware of its presence because of its gravitational pull, which is essential to comprehending everything from the birth of galaxies such as the Milky Way, of which we are a part, to the large-scale development of the universe. Its nature, however, is largely unknown and represents one of the biggest unresolved modern physics problems. Recently, the fuzzy DM hypothesis has been investigated as a possible option. According to this hypothesis, DM is made up of particles known as axions, which act like waves because of their extremely low mass. Because photons have wavelengths on galactic sizes of millions of light years, they behave similarly to light on much larger scales. These particles resolve important cold DM model differences by forming dense cores in galactic centers and suppressing small-scale features \cite{banares2023confronting,zupancic2024fuzzy,nori2021scaling}.

According to Einasto \cite{einasto1969galactic}, a model must meet certain requirements for its descriptive capabilities in order to faithfully represent real-world galactic systems. Since the density profile is the source from which the surface mass density, cumulative mass profile, and gravitational potential are derived, it makes sense to choose it as the descriptive function first. The density profile of a physical model must be finite, non-negative, and gradually falling to zero at bigger radii, among other conditions. Additionally, some moments of the mass function, such as those describing the system's total mass, effective radius, and central gravitational potential, have to be finite, and the description functions cannot have jump gaps (or discontinuities). A number of legitimate families of descriptive functions, such as the Einasto density profile (EDP), were developed by the Einasto \cite{einasto1969galactic}. Einasto \cite{einasto1969andromeda} used the EDP to model M31. Einasto \cite{einasto1972galactic} then applied the model to some surrounding galaxies, including M87, M32, the Milky Way, the Sculptor dwarfs and Fornax, and M31. Each component of these models represented a different physically homogenous population of stars, each having its own set of parameters ${n, h}$, and $\mathcal{P}_0$, which stood for Einasto index, scale length, and central density, respectively. The formulation of EDP can be utilized as a viable approach for constructing fuzzy dark matter WH solutions.

Herrera and collaborators explored the key consequences of the gravitational collapse within the realms of the Israel-Stewart formalism, focusing on the analysis of viscous dissipation involving bulk shear viscosity \cite{herrera2009dynamics}. In the context of postquasi-time independent approximations, Herrera investigated the solutions for a relativistic self-gravitational collapse in environments characterized by dissipative energy effects \cite{herrera2002relativistic}. Odintsov and Oikonomou \cite{odintsov2021neutron} investigated the fluid dynamics of rotating celestial bodies and established unique connections between inflationary attractors and the characteristics of compact stars to achieve stable configurations \cite{oikonomou2024phenomenology}. Nojiri \emph{et al.} \cite{nojiri2024propagation} introduced additional gravitational interaction terms to study the transmission of gravitational waves. They aimed to propose a framework where spherical relativistic entities could generate gravitational wave velocities equivalent to the light speed in a vacuum environment. Building upon the context provided, our exploration delves into the potential application of fuzzy WHs in characterizing compact self-gravitating structures. Our focus centers on utilizing the Einasto-density profile (EDP) model to describe dark matter (DM) haloes, facilitating the development of a feasible interaction between WHs and DM haloes. Specifically, we introduce the Einasto-density profile as a foundational framework for our theoretical representation of the fuzzy WH phenomenon. This investigation aims to bridge the realms of theoretical physics and astrophysical observations, offering a deep understanding of the intriguing interaction between WHs and DM distributions within the cosmos \cite{merritt2006empirical}.

Furthermore, Herrera \emph{et al}. \cite{herrera2004spherically} determined that the system of equations that brings about the true scenario of disappearing spatial gradients of energy density is a collapsing source that collapses self-gravitating and has an anisotropic matter distribution. Additionally, the complexity factor of astrophysical configuration under modified gravity has been discussed by Yousaf and coworkers in various contexts \cite{yousaf2022f}. The main goal of this work is to introduce a theoretical model of a time-independent, spherical symmetry WH based on FDM using the extended theories of gravity. The primary differences between FDM WHs and regular WHs are their stability criteria, interactions with nearby DM, structural complexity, and possible astrophysical effects. The WH structures become more attainable and significant in contemporary cosmology research when fuzzy DM is incorporated, suggesting a more intricate relationship between unusual structures and the universe's fabric. A way to distinguish between $f(R)$ or EGB WHs, BHs, and neutron stars is provided by the Einasto profile, which also affects lensing, accretion dynamics, and structure. WHs differ from BHs in gravitational lensing because of the smooth dark matter distribution, which modifies light bending. The profile influences the way matter interacts with these objects in accretion dynamics, which impacts the stability of WHs and the flow of material surrounding BHs. Additionally, it affects neutron stars by changing the distribution of DM, which can modify their visible radiation and mass-radius relationship. The profile affects internal composition and stability in altered gravity models such as EGB or $f(R)$, offering important information for identifying these items and evaluating altered gravity theories \cite{nojiri2024wormholes}.

 The manuscript is structured as follows: Section II introduces the basic framework of field equations along with essential definitions. Section III explores violations of the NEC. Section IV offers a concise overview of the EDP. Section V delves into conservation equations. Section VI discusses the impact of active gravitational mass, while section VII addresses the complexity factor. The final section presents the conclusions and a summary of the discussion.

\section{Basic Formalism for Wormhole Models in Extended Gravity}

In this section, we provide the mathematical formalism to describe WH models in two well-known extended theories of gravity. First, we describe our theoretical modeling in the 4D EGB gravity model and then we present the equations related to $f(R)$ gravity.

\subsection{4D EGB model of gravity}

This section will provide a concise introduction to the 4D EGB model of gravity. Our formulation follows Glavan and Lin’s prescription for 4D EGB gravity \cite{glavan2020einstein}, where the GB coupling is rescaled as $\alpha \times (D-4)^{-1}$ and the limit $D\rightarrow4$ is taken after variation. Although this method introduces well-behaved field equations in 4D via dimensional regularization, its physical viability has been questioned \cite{gurses2020there}. We treat this as an effective model capturing higher-curvature effects in 4D spacetime. The action for the EGB background is defined as \cite{glavan2020einstein}
\begin{align}\label{kbb1}
I_{\mathcal{G}}=\int \sqrt{-g} \left(\frac{\alpha}{D-4} \mathfrak{L}_{\mathcal{G}\mathcal{B}}+L_m+R\right) \, d^{D}x,
\end{align}
where $L_m$ stands of matter Lagrangian, the variable $g$ signifies the determinant of $g_{\mu \nu}$, while $\alpha$ stands for the GB coupling coefficient. This paper will present the discussion related to the scenario where $\alpha \geq 0$. The Lagrangian term $\mathfrak{L}_{\mathcal{G}\mathcal{B}}$ is defined as
\begin{align}\label{a1}
\mathfrak{L}_{\mathcal{G}\mathcal{B}}= -4R^{\tau\eta}R_{\tau\eta}+R^2+R^{\tau\eta\alpha\beta} R_{\tau\eta\alpha\beta}.
\end{align}
On applying the variational technique concerning $g_{\tau\eta}$ on the equation described above, the following equation of motion results
\begin{align}\label{a2}
G_{\tau\eta}+\frac{\alpha}{D-4} \mathcal{S}_{\tau\eta} = \mathcal {K} T_{\tau\eta},
\end{align}
where $T_{\tau\eta}$ is stress-energy momentum tensor and is defined as $T_{\tau\eta}=\dfrac{-1}{\sqrt{-g}}\dfrac{\delta(\sqrt{-g} L_m)}{\delta g^{\tau\eta}}$ and
\begin{align}\label{kbb2}
G_{\tau\eta}=R_{\tau\eta}-\frac{1}{2} {R}{g_{\tau\eta}},
\end{align}
\begin{align}\label{kbb3}
 \mathcal{S}_{\tau\eta}=4(\frac{1}{2}RR_{\tau\eta}-R_{\tau\alpha}R^{\alpha}_{\eta}-R_{\tau\alpha\eta\beta}R^{\alpha\beta}
 -\frac{1}{2}R_{\tau\alpha\beta\gamma}R^{\alpha\beta\gamma}~_{\eta}-\frac{1}{8} \mathfrak{L}_{\mathcal{G}\mathcal{B}} g_{\tau\eta}).
\end{align}
In the realms of 4D spacetime, the GB term is typically associated with the total derivative, leading to no discernible impact on the field equations. Yet, when adjusting the coupling parameter to $\frac{\alpha}{D-4}$, the GB term musto for $D=4$, allows for further exploration of its contribution.

This article aims to investigate how anisotropic pressure influences the presence of fuzzy WH. The approach involves considering relativistic anisotropic matter sources, given as follows
\begin{align}\label{a3}
T_{\tau\eta}= ({\rho}+{P_t}){\upsilon_{\tau}}{\upsilon_{\eta}}-{P_t}{g_{\tau\eta}} +{\varrho}{\varkappa_{\tau}}{\varkappa_{\eta}},
\end{align}
where ${\rho}$ stands for the fluid density, the variable ${\varrho = P_r - P_t}$, ${P_t}$ and ${P_r}$ are the tangential and radial pressure components respectively. Additionally, the velocity-four vector ${\upsilon_{\tau}}$ and the radial unit-four vector ${\varkappa_{\tau}}$ adhere to the following conditions within the co-moving coordinate system: ${\upsilon^{\tau}}{\upsilon_{\tau}}=1$ and ${\varkappa^{\tau}}{\varkappa_{\tau}}=-1$.

We now delve into the subject of fuzzy WHs with spherical symmetry. Our initial emphasis will be on examining the static spherically symmetric geometry \cite{morris1988wormholes}
\begin{align}\label{a4}
ds^2=g_{\tau \eta} dx^\tau dx^\eta = \exp[2a(r)]dt^{2}-\left({1-\frac{b(r)}{r}}\right)^{-1}dr^{2}-r^2d\Omega^2.
\end{align}
In the above expression $d\Omega^2=d\theta^{2} +\sin^2\theta{d\phi^2}$, the function $a(r)=a$ signifies a radial red-shift function, while $b(r)=b$ represents the shape function describing the geometry of the WH. In equation \eqref{a3}, the quantities ${\upsilon^{\tau}}$ and ${\varkappa^{\tau}}$ are defined as ${\upsilon^{\tau}}=\exp[-a(r)/2] \delta^{\tau}_{0}$ and ${\varkappa^{\tau}}=-\dfrac{1}{\sqrt{1-\frac{b(r)}{r}}} \delta^{\tau}_{1}$, respectively. To create stable surface structures resembling a WH's throat, it is essential to adjust the radial distance from a defined point $r_0$ to infinity. This modification is necessary to ensure the throat expands correctly, meeting the required flare-out conditions. Specifically, these conditions can be provided as $\dfrac{b(r)-rb'(r)}{b(r)^2}>0$, with $b'(r_0) < 1$. Within the EGB model framework, it is noteworthy that these limitations result in the creation of a WH associated with exotic matter, thereby breaching the NEC.

In the scenario where the $D\rightarrow 4$, the quantities associated with the matter are explicitly represented as
\begin{align}\label{a5}
\rho &= \frac{1}{r^6} \left\{{\alpha  b \left[2 r b'-3 b\right]}+{r^4 b'}\right\},
\\\label{a6}
P_ {\mathit {r}} &= \frac{1}{r^6} \left\{{\alpha  b \left[4 r (r-b) a'+b\right]}+[2 r^4 (r-b) a']-{b}{r^3}\right\},
\\\nonumber
P_ {\mathit {t}}&=\frac{1}{2r^6} \{r a'(12\alpha b+r^3-8r\alpha)(b-rb')+[r^3-2b\alpha][2r(r-b)
\\\label{a7}&\quad\quad
a'-r b'] +2r^2 [r-b][4 b\alpha +r^3][a''+a'^2]-4\alpha b^2\}.
\end{align}
Equations \eqref{a5} \textendash \eqref{a7} represent the field equations involving five unknown variables: $\rho$, $P_r$, $P_t$, $a$, and $b$. There exist numerous techniques available for solving such equations. Here, we outline the suggested approach to tackle these field equations.

\subsection{Field Equations in $f(R)$ Gravity}

A particular alternative form of GR is the $f(R)$ model of gravity. This altered model of gravity encompasses various gravitational theories, each distinguished by a distinct Ricci-scalar function. The most basic rendition of this function is when it mirrors the scalar itself, effectively reverting to GR. The inclusion of a customizable function within $f(R)$ gravity seeks to offer versatility in elucidating the universe's structural formation and tackling the issue of late-time cosmic acceleration. Buchdahl \cite{buchdahl1970non} first proposed this concept, using $f$ as a placeholder for the arbitrary function, and Starobinsky \cite{starobinsky1980new} extensively studied cosmic inflation using this gravity theory. This article examines the WH geometry in the $f(R)$ model. One can deduce the field equations governing $f(R)$ model as
\begin{align}\label{ia2}
(R_{\xi\psi}-\nabla_{\xi} \nabla_{\psi} +g_{\xi\psi}\Box)F-\frac{1}{2} f g_{\xi\psi}=T_{\xi\psi}.
\end{align}
The function $f$'s derivative with respect to the $R$ is represented as $F$. By contracting Eq. \eqref{a2}, we can derive a new expression as
\begin{align}\label{ia3}
3\Box F+RF-2f= T.
\end{align}
By considering the trace of the $T^{\xi\psi}$ as $T = g_{\xi\psi}T^{\xi\psi}$ and performing the necessary contractions and rearrangements, one can derive the field equations relative to $f(R)$ model as
\begin{align}\label{ia4}
  G_{\xi\psi}=R_{\xi\psi}-\frac{1}{2} R g_{\xi\psi}=T_{\xi\psi} ^{ef}.
\end{align}
The right side of Eq. \eqref{ia4} corresponds to the stress energy-momentum tensor, $T_{\xi\psi} ^{ef}$, which is obtained from the curvature stress energy-momentum tensor, $T_{\xi\psi} ^{c}=\frac{1}{F} \left(\nabla_{\xi} \nabla_{\psi} - \frac{1}{4}(RF+\Box F + T) g_{\xi\psi} \right)$ and $\hat{T}_{\xi\psi} ^{(m)}=\frac{T_{\xi\psi}^{(m)}}{F}$, the stress-energy momentum tensor for matter is provided as $T_{\xi\psi}^{(m)}$. The tensor, which serves to describe the matter content within a WH, can be expressed as an anisotropic matter's distribution given in Eq. \eqref{a3}. The focus is on the $f(R)$ gravity model in the metric formalism, with metric coefficients that are not dependent on time. We use Morris-Thorne spacetime to find exact solutions for traversable wormholes
without violating energy and pressure conditions. We propose the power-law form of $f(R)$, where $f(R) = f_0 R^{1+\lambda}$, and $\lambda$
is a real number. By setting $\lambda$ to zero, GR can be recovered. The study of $f(R)$ theories of gravity
was prompted by the necessity to characterize the post and pre-cosmic histories of our universe. The latest cosmic observational
data on the acceleration-deceleration transition of the post-universe served as the primary impetus for investigating these theories.
This stipulation placed limitations on $f(R)$ models, enabling feasible options for $f(R)$ models. Such theoretical frameworks
evade the Ostrogradski instability and do not include involvement from invariants of curvature besides $R$
\cite{lanczos1938remarkable,zwiebach1985curvature,de2010f,cognola2008class,li2007cosmology,nojiri2007unifying}.
In this context, the energy density of matter is symbolized by the variable $\rho$. Meanwhile, the pressures experienced in the transverse and radial directions are represented as $P_t$ and $P_r$, respectively. With a constant red-shift function, the field equations associated with $f(R)$ model of gravity are provided as
\begin{align}\label{ia6}
\rho &=\frac{1}{r^2} \left[{F b'}\right],
\\\label{ia7}
P_r&= \frac{1}{2r^3}\left\{{F' \left[r b'-b\right]r}+{2F''[b-r]r^2}-2{b F}\right\},
\\\label{ia8}
P_t&=\frac{1}{2 r^3}\left\{{F \left(b-r b'\right)}+{2F'r[b-r]}\right\}.
\end{align}
The interior of the celestial bodies of the cosmos, like galaxies and clusters of galaxies, undergoes evolution within a dynamic medium, indicating a non-linear phase. To comprehend the formation of their structure, it is essential to study their linear, quasilinear, or nearly linear phases. Analyzing the non-linear gravitational interactions by using traditional mathematical approaches is often challenging. Researchers have employed different assumptions and numerical methods to address this complexity. In this context, we will focus on a specific category of the $f(R)$ model, firstly introduced by Starobinsky \cite{starobinsky1980new}
\begin{align}\label{ia9}
f(R)=R+\alpha R^2.
\end{align}
The model in the above expression contains a parameter, denoted as $\alpha$, which can take on any real value. When this parameter is assigned a value of zero, the model corresponds to the dynamics of GR. To obtain the specific solutions for this model, we are required to solve the field equations for this particular case and then simplify the results accordingly
\begin{align}\label{ia10}
\rho &=\frac{1}{r^2} \left\{{[2 \alpha  R+1] b'}\right\},
\\\label{ia11}
P_r&=\frac{1}{r^3}\left\{{\alpha \left[r b'-b\right]r}-{b [2 \alpha  R+1]}\right\},
\\\label{ia12}
P_t&=\frac{1}{2r^3}\left\{{[2 \alpha  R+1] \left[b-r b'\right]}+{4 \alpha  [b-r]}\right\}.
\end{align}
We chose $f(R)=R+\alpha R^2$ due to its analytical tractability and historical relevance. However, our framework can be generalized to include viable cosmological models such as Hu–Sawicki or exponential gravity models, which will be explored numerically in future studies \cite{hu2007models}.

\section{Wormhole Solution}

It is widely acknowledged that energy conditions are violated in time-independent, spherically symmetric, four-dimensional space-time setups. This breach of ECs is a significant occurrence within this particular space-time context. The flaring-out conditions dictate the presence of these violations. Additionally, in theories involving higher dimensions, it is conceivable that the regions around the throat of a WH are the only locations where EC violations can be avoided or satisfied. These conditions in extended gravity can be expressed as follows
\begin{align}\label{a8}
T_{\xi\sigma} \kappa^{\xi} \kappa^{\sigma} \geq 0,
\end{align}
when $\kappa$ denotes a null vector. The NEC can be expressed for an anisotropic fluid as
\begin{align}\nonumber
 \rho + P_ {r} \geq 0  ~~and~~   \rho + P_ {t} \geq 0.
\end{align}
The equations \eqref{a5} \textendash  \eqref{a7} provide the following expressions for NEC as
\begin{align}\label{a9}
\rho+P_ {r} &=\frac{1}{r^6} \left\{{\left[2 \alpha  b+r^3\right] \left[r \left(2 r a'+b'\right)-b \left(2 r a'+1\right)\right]}\right\},
\\\nonumber
\rho+P_{\mathit{t}}&=\frac{1}{2 r^6}\{r b [r^2 (1-2 (r^2-4 \alpha ) a'')-r a' (12 \alpha +12 \alpha  b'+r^2)
\\\nonumber&\quad
-2 (r^4-4 \alpha  r^2) a'^2+6 \alpha  b']+r^3 [r (2 r^2 a''+b')+a' (2 r^2-
\\\label{a10}&\quad
(r^2-8 \alpha ) b')+2 r^3 a'^2]-4 \alpha  b^2 [2 r^2 a''+2 r^2 a'^2-4 r a'+3]\}.
\end{align}
The aforementioned equations show that the flare-out condition causes the NEC to be violated when $a'(r)=0$. We adopt a constant redshift function for tractability and to isolate the geometric contributions of shape function and density. This choice ensures regularity and avoids event horizons. A variable redshift function could model time-dependent effects that would require numerics, which is beyond our current scope. Equation \eqref{a9} transforms at the throat of the WH (where $r=r_0$) as
\begin{align}\label{a11}
\rho(r_0)+P_ {r}(r_0) =\frac{1}{r_0^6} \left\{{\left[2 \alpha  b(r_0)+r_0^3\right] \left[r_0 \left(2 r_0 a'(r_0)+b'(r_0)\right)-b(r_0) \left(2 r_0 a'(r_0)+1\right)\right]}\right\}.
\end{align}

By making use of Eqs. \eqref{ia10} \textendash \eqref{ia12}, we get NEC expression in $f(R)$ gravity are as under
\begin{align}\label{ia14}
\rho+P_ {r}|_{f(R)}&=\frac{\left(r b'-b\right) (\alpha  r+2 \alpha  R+1)}{r^3},
\\\label{ia15}
\rho+P_{t}|_{f(R)}&=\frac{r \left((2 \alpha  R+1) b'-4 \alpha  r\right)+b (4 \alpha  r+2 \alpha  R+1)}{2 r^3},
\end{align}
where the notation $``|_{f(R)}"$ indicates that the quantity is calculated in $f(R)$ gravity. The given equations suggest that the flaring-out condition outcomes in the breach of NEC. Specifically, Eq. \eqref{ia14} transforms the WH throat, where the radius $r$ equals the throat radius $r_0$
\begin{align}\label{ia16}
\rho(r_0)+P_ {r}(r_0)|_{f(R)} =\frac{\left[r_0 b'(r_0)-b(r_0)\right] [\alpha  r_0+2 \alpha  R+1]}{r_0^3}.
\end{align}

\section{EDP Model}

Einasto profile is a practical and accurate way to model the distribution of matter in regions where WHs could be present, particularly in environments where DM is more prevalent. Its smooth, restricted nature and empirical success in defining astrophysical formations make it a valuable tool for theoretical investigation into WH solutions \cite{batic2021fuzzy}. The latest work emphasizes the connection between the Einasto profile and the WH structure, particularly in the context of alternative gravity theory \cite{naseer2024imprints}. Although Newtonian hydrodynamics is effective in describing the bulk behavior of CDM at galactic scales, the core regions, particularly within $\sim 1 kpc$, exhibit behaviors better captured by quantum and relativistic effects \cite{hui2017ultralight}. Our use of a relativistic framework is motivated by the need to capture such effects, especially where the $n$ is low ($n \leq0.5$), corresponding to FDM-dominated bulges.

To enhance our understanding of the mysterious components present in the cosmos, such as DM haloes surrounding galaxy clusters, it is essential to explore their characteristics more profoundly. The studies involving cosmological simulations that analyze the development of structures through N-body techniques reveal that distinct density models with three specific parameters can efficiently describe different galaxy clusters or DM haloes \cite{merritt2006empirical,hayashi2008understanding,gao2008redshift,yousaf2023generating}. A model that is commonly used for DM haloes is the EDP model, a 3D adaptation of the S\'{e}rsic model \cite{gao2008redshift,navarro2010diversity,de2019estimation}. This model is utilized in analyzing the brightness of ancient cosmic structures and the central regions of spiral galaxies \cite{gadotti2009structural}. The EDP model, with its three parameters, provides a more accurate representation of DM haloes compared to some special forms of the halo model, which is characterized by only two parameters \cite{navarro1997universal}. The EDP displays a power-law format with a logarithmic slope \cite{retana2012analytical,baes2022einasto} as
 \begin{align}\label{a12}
\wp (r)=\frac{d \ln (\rho )}{d \ln (r)}(r)\propto r^{\frac{1}{n}},
\end{align}
here $n$ is known as the Einasto index and this parameter helps in defining some particular form of the EDP model. One can obtain a general density profile through the integration of this parameter \cite{de2019estimation} as
\begin{align}\label{a13}
\ln \left(\frac{\rho (r)}{\rho _m}\right)=-c_n\left(\left(\frac{r}{r_m}\right){}^{\frac{1}{n}}-1\right).
\end{align}
In this realm, the symbol ${\mathit{r_m}}$ represents the radius of the sphere containing half of the total mass, while ${\mathit{c_n}}$ is a parameter that influences ${\mathit{r_m}}$, with ${ \rho _m= \rho _m (r) \text{and } \rho _0 =\rho _m e^{c_n}}$. The value of ${\mathit{c_n}}$ is a fixed numerical value that ensures ${\rho _m}$ accurately describes the radius enclosing half of the mass. It is essential to note that multiple parameterizations of the EDP exist, each introducing distinct sets of independent parameters. Within the domain of DM-haloes, a commonly adopted representation follows a specific structure as
\begin{align}\label{a14}
\ln \left(\frac{\rho (r)}{\rho _{-2}}\right)=-2 n\left(\left(\frac{r}{r_{-2}}\right){}^{\frac{1}{n}}-1\right).
\end{align}
Here $r_{-2}$ and $\rho_{-2}$ denote the radius and density, respectively, while the density function $\rho(r)$ is directly proportional to $r^{-2}$. One can write
\begin{align}\label{a15}
\rho (r)=\rho _0 \exp{[-\left(\frac{r}{h}\right)^\frac{1}{n}]},
\end{align}
while the scale length $h$ is specified as
\begin{align}\label{a16}
h=\frac{r_m}{c_n^n}=\frac{r_{-2}}{(2 n)^n},
\end{align}
and the central density is provided as
\begin{align}\label{a17}
\rho _0=\rho _m e^{c_n}=\rho _{-2} e^{2 n}.
\end{align}
The descriptive components of the model need to satisfy particular standards such as gravitational potential, mass profile, and surface mass density to faithfully depict actual galactic formations \cite{einasto1969galactic}. The selection of these descriptive functions forms the basis for accurately representing any galactic structure model. Since the density profile plays a crucial role in each of these functions, it emerges as the most advantageous option in this context. For any value of $r$, the function $\rho(r)$ must remain positive and finite. It should exhibit a slight decrease, getting closer to zero as $r$ moves farther and closer to infinity. The density functions ought to link with a system that exhibits total mass, finite multi-pole expansion, and effective radius. Smooth transitions should be displayed by these descriptive functions; jump discontinuities and other abrupt or sudden shifts should be avoided. Any model aiming to faithfully depict real-world galactic structure must fulfill these criteria. The Einasto index $n$ controls the steepness of the density profile. Low $n$ corresponds to FDM-like cored profiles, enabling stable WH throats. As $n$ rises, the density flattens, weakening the pressure gradients needed to satisfy flaring-out and equilibrium conditions. Thus, WH viability is more pronounced in low $n$ environments, consistent with FDM cores. For FDM particles of mass $m \sim 10^{-23}$ eV and central densities $\rho_0 \sim 10^{-24}g/cm^3$, the typical throat radius lies in the $0.1-1 kpc$ range, overlapping with observed halo cores. This supports the plausibility of wormhole-like geometries forming in dense galactic nuclei. These parameters are compatible with galaxy bulges and cores such as M87, M32, M31, Fornax, Sculptor, and the Milky Way, supporting the astrophysical feasibility of such structures forming in FDM-dominated halos. Our model primarily focuses on FDM through the Einasto profile, we recognize that galactic halos are likely composed of both CDM and FDM components. This hybrid nature has been suggested in structure formation simulations \cite{de2019estimation}. In this work, the EDP serves as an effective tool to capture both cored and cuspy behaviors depending on the choice of Einasto index $n$, thereby implicitly accommodating hybrid DM distributions. In the following, we shall calculate the value of the shape function that led to the form of the equations of state parameters in the subsequent theories.

\subsection{Calculations of shape function and EOS in 4D EGB Gravity}

In GR, the traversable WH are believed to breach the NEC due to their exotic matter content. In the realms of modified gravity, breaching of ECs often arise from the effective geometric terms in the field equations rather than the physical matter sector. Thus, NEC violation does not imply exotic matter in the traditional GR sense but reflects contributions through higher-order curvature effects \cite{capozziello2010dark,harko2013modified,ilyas2023traversable}. A sustainable WH may be formed through a gradual collapse process. Another feasible approach is to assess the stability of a traversable WH with a sizable throat. The determination of $b(r)$ in 4D EGB gravity involves the analysis of Eqs. \eqref{a5} and \eqref{a15}, leading to the derivation of the differential equation (DE) in a specific format as
\begin{align}\label{a18}
\frac{\alpha  b \left[2 r b'-3 b\right]}{r^6}+\frac{b'}{r^2}=\rho_0 \exp{[-\left(\frac{r}{\mathit{h}}\right)^{1/\mathit{n}}]}.
\end{align}
The function $b$ can be derived by solving the provided DE and the integration constant ($c$) in the above equation can be found by using $b({r}_0)={r}_0$ as under
\begin{align}\label{a20}
\mathit{c}=\frac{\alpha +\mathit{h}^3 \mathit{n} \rho_0 \mathit{r}_0 \Gamma \left(3 \mathit{n},\left(\frac{\mathit{r}_0}{\mathit{h}}\right){}^{1/\mathit{n}}\right)+\mathit{r}_0^2}{\alpha  \mathit{r}_0}.
\end{align}
Thus $b(r)$ in 4D EGB gravity takes the form
\begin{align}\label{a21}
b(r)=\frac{\pm\sqrt{\alpha } r \sqrt{r \left[\frac{r^3}{\alpha }+\frac{4 \alpha }{\mathit{r}_0}+4 \mathit{h}^3 \mathit{n} \rho_0 \left[\Gamma \left(3 \mathit{n},\left(\frac{\mathit{r}_0}{\mathit{h}}\right){}^{1/\mathit{n}}\right)-\Gamma \left(3 \mathit{n},\left(\frac{r}{\mathit{h}}\right)^{1/\mathit{n}}\right)\right]+4 \mathit{r}_0\right]}-r^3}{2 \alpha}.
\end{align}
Although Eq. (\ref{a21}) is analytically involved, its near-throat behavior can be approximated via a power-series expansion in $r$, capturing the geometry’s leading features. These approximations are consistent with observationally supported galactic core sizes ($\sim100-1000 pc$) and permit physically meaningful insights into curvature and flare-out conditions. In this equation, there are two shape functions associated with the ($\pm$) signs. We have verified the condition of asymptotic flatness for both functions, and the shape function with a positive sign meets the required criteria, i.e., $\frac{b(r)}{r}\rightarrow0$ as $r\rightarrow\infty$. The dynamics of the shape function are illustrated in Figs. \textbf{\ref{1f}} and \textbf{\ref{2f}} in Appendix B. In Fig. \textbf{\ref{1f}} (mentioned in Appendix B), the left graph demonstrates that at $r_0=0.5$, $b(0.5)=0.5$. The right graph in Fig. \textbf{\ref{1f}} in Appendix B confirms the satisfaction of flaring-out conditions. Furthermore, Fig. \textbf{\ref{2f}} shows asymptotic flatness in the left graph. The right graph in Fig. \textbf{\ref{2f}} in Appendix B indicates the WH throat location at $r_0 = 0.5$, where the curve $b(r) - r$ intersects the $r$-axis.

To gain insight into the behavior of the GB parameter $\alpha$, we have illustrated the characteristics of WH conditions using contour plots as depicted in Fig. \textbf{\ref{3f}} (mentioned in Appendix B). The analysis of Fig. \textbf{\ref{3f}} in Appendix B reveals that as $\alpha$ rises, the behavior of $b(r)$ intensifies in the vicinity of the WH throat. This indicates a clear correlation between $\alpha$ and the shape function near the WH throat. Furthermore, the right plot of Fig. \textbf{\ref{3f}} in Appendix B demonstrates that as $\alpha$ increases, the derivative of $b(r)$ is less than $1$. For the previously derived shape function, the WH metric can be expressed as
\begin{align}\label{a22}
ds^2=dt^{2}-(\frac{2 \alpha -\beta_1+r^2}{2 \alpha })^{-1}dr^{2}.
-r^2(d\theta^{2} +\sin^2\theta{d\phi^2}).
\end{align}
When we consider $a'(r)=0$, the resulting values of $P_r$ and $P_t$ in 4D EGB gravity will become
\begin{align}\label{a23}
P_r &=\frac{r^3+\frac{\alpha ^2}{\mathit{r}_0}+\alpha  \mathit{r}_0-\beta_1  r+\alpha  \mathit{h}^3 \mathit{n} \rho_0 \left[\Gamma \left(3 \mathit{n},\left(\frac{\mathit{r}_0}{\mathit{h}}\right){}^{1/\mathit{n}}\right)-\Gamma \left(3 \mathit{n},\left(\frac{r}{\mathit{h}}\right)^{1/\mathit{n}}\right)\right]}{\alpha  r^3},
\\\nonumber
P_t&=\frac{1}{2 \alpha ^3 \beta_1  r^3 r_0} ((2 r^2-3 \beta_1 ) (\alpha ^2+\alpha \rho_0 h^3 n (-r_0) [\Gamma \left(3 n,(\frac{r}{h})^{1/n}\right)
\\\label{a24}&
-\Gamma\left(3 {n},(\frac{{r}_0}{{h}}){}^{1/{n}}\right)]+2 {r}_0^2)+\alpha \rho_0 {r}_0 r^3 (\beta_1 -2 r^2) e^{(\frac{r}{{h}})^{1/{n}}}-{r}_0 r^3 (r^2-\beta_1 )),
\end{align}
where
\begin{align}\label{a25}
\beta_1=\sqrt{\alpha } \sqrt{r \left[\frac{r^3}{\alpha }+\frac{4 \alpha }{\mathit{r}_0}+4 \mathit{h}^3 \mathit{n} \rho_0 \left[\Gamma \left(3 \mathit{n},\left(\frac{\mathit{r}_0}{\mathit{h}}\right){}^{1/\mathit{n}}\right)-\Gamma \left(3 \mathit{n},\left(\frac{r}{\mathit{h}}\right)^{1/\mathit{n}}\right)\right]+4 \mathit{r}_0\right]}.
\end{align}
The equations of state (EOS) for the tangential and radial directions are given by the following expressions
\begin{align}\label{eoseq}
  \mathcal{W}_r = \frac{P_1}{\rho}, \quad  \mathcal{W}_t = \frac{P_2}{\rho}.
\end{align}
These equations represent the relationships between pressure, density, and the respective directions in the system and have the following mathematical representation in 4D EGB gravity as
\begin{align}\label{a26}
\mathcal{W}_r=\frac{\exp{[\left(\frac{r}{\mathit{h}}\right)^{1/\mathit{n}}]} \left\{r^3+\frac{\alpha ^2}{\mathit{r}_0}+\alpha  \mathit{r}_0-\beta_1  r+\alpha  \mathit{h}^3 \mathit{n} \rho_0 \left[\Gamma \left(3 \mathit{n},\left(\frac{\mathit{r}_0}{\mathit{h}}\right){}^{1/\mathit{n}}\right)-\Gamma \left(3 \mathit{n},\left(\frac{r}{\mathit{h}}\right)^{1/\mathit{n}}\right)\right]\right\}}{\alpha  \rho_0 r^3},
\end{align}
\begin{align}\nonumber
\mathcal{W}_t&=\frac{\exp{[\left(\frac{r}{\mathit{h}}\right)^{1/\mathit{n}}]}}{2 \rho_0  \alpha ^3 \beta_1  r^3 r_0} ((2 r^2-3 \beta_1 ) (\alpha ^2+\alpha \rho_0 h^3 n (-r_0) [\Gamma \left(3 n,(\frac{r}{h})^{1/n}\right)
\\\label{a27}&
-\Gamma\left(3 {n},(\frac{{r}_0}{{h}}){}^{1/{n}}\right)]+2 {r}_0^2)+\alpha \rho_0 {r}_0 r^3 (\beta_1 -2 r^2) \exp{[(\frac{r}{{h}})^{1/{n}}]}-{r}_0 r^3 (r^2-\beta_1 )),
\end{align}
 \begin{figure}[t]
{{\includegraphics[height=2.5 in, width=3.0 in]{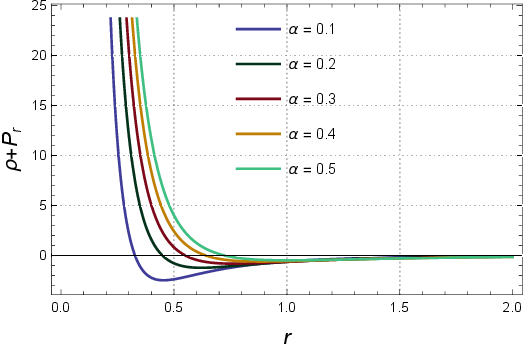}}}
{{\includegraphics[height=2.5 in, width=3.0 in]{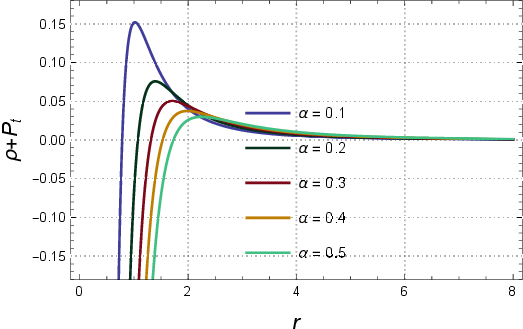}}}
{{\includegraphics[height=2.5 in, width=3.0 in]{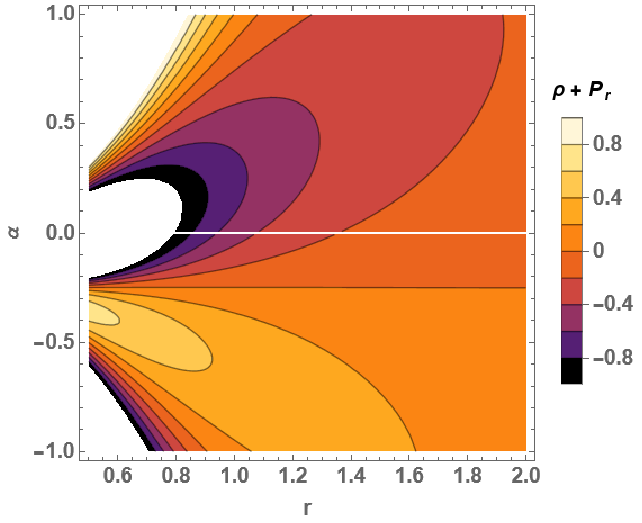}}}
{{\includegraphics[height=2.5 in, width=3.0 in]{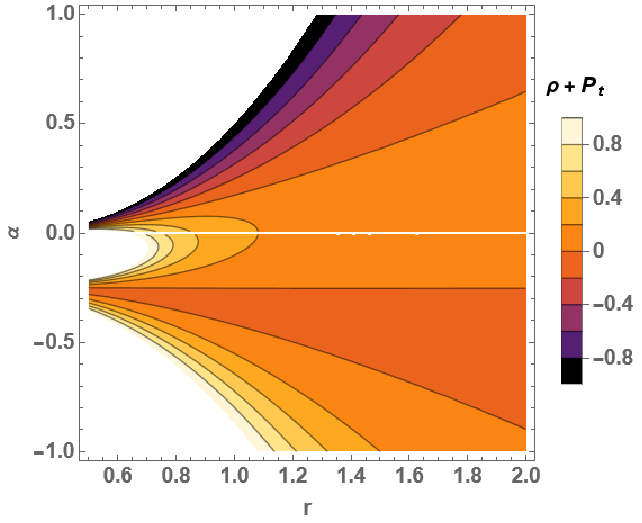}}}
\caption{The dynamics of $\rho+P_r$ and $\rho+P_t$ versus radial axis $r$ corresponding to various $\alpha$ in 4D EGB theory.}\label{5f}
\end{figure}
Figure \textbf{\ref{4f}} in Appendix B displays the graphs of $\mathcal{W}_{\mathit{r}}$ and $\mathcal{W}_{\mathit{t}}$ along radial axis $r$. The behavior of $\mathcal{W}_{\mathit{t}}$ is depicted on the left side of Fig. \textbf{\ref{4f}} in Appendix B. As $\mathcal{W}_{\mathit{t}}$ approaches the WH-throat, it experiences a decrease, followed by an increase as it moves away from this location. The regions where the NEC conditions are satisfied, specifically when $\rho+P_r\geq0$ and $\rho+P_t\geq0$, have been identified and are visualized in Fig. \textbf{\ref{5f}}.

\subsection{Calculations of shape function and EOS in $f(R)$ Gravity}

The calculation of $b(r)$ requires the examination of equations \eqref{ia10} and \eqref{ia15}, which ultimately results in the development of a particular differential equation of first order in $f(R)$ gravity as
\begin{align}\label{ja18}
\frac{(2 \alpha  R+1) b'(r)}{r^2}=\exp{\left[\left(-\left(\frac{r}{h}\right)^\frac{1}{n}\right)\right]} \mathcal{P}_0.
\end{align}
We consider the present values of the Ricci curvature scalar (i.e., $R=\widetilde{R}$) and solve the above equation to find $b(r)$. The examination of constant curvature solutions within $f(R)$ model of gravity was carried out by Odintsov and his associates and some researchers, as described in reference \cite{cognola2005one,sharif2013dynamical}
\begin{align}\label{ja19}
b(r)|_{f(R)}=d-\frac{\mathit{h}^3 \mathit{n} \mathcal{P}_0 \Gamma \left(3 \mathit{n},\left(\frac{r}{\mathit{h}}\right)^{1/\mathit{n}}\right)}{2 \alpha  \widetilde{R}+1}.
\end{align}
The integration constant $d$, can be found by using the throat condition, i.e.,
\begin{align}\nonumber
b({r}_0)={r}_0,
\end{align}
which eventually yields in $f(R)$ gravity as
\begin{align}\label{ja20}
d=\frac{\mathit{h}^3 \mathit{n} \mathcal{P}_0 \Gamma \left(3 \mathit{n},\left(\frac{\mathit{r}_0}{\mathit{h}}\right){}^{1/\mathit{n}}\right)+\mathit{r}_0+2 \alpha  \mathit{r}_0 \widetilde{R}}{2 \alpha  \widetilde{R}+1},
\end{align}
so the shape function will in $f(R)$ gravity become
\begin{align}\label{ja21}
b(r)|_{f(R)}=\frac{\mathit{h}^3 \mathit{n} \mathcal{P}_0 \left[\Gamma \left(3 \mathit{n},\left(\frac{\mathit{r}_0}{\mathit{h}}\right){}^{1/\mathit{n}}\right)+\Gamma \left(3 \mathit{n},\left(\frac{r}{\mathit{h}}\right)^{1/\mathit{n}}\right)\right]}{2 \alpha  \widetilde{R}+1}+\mathit{r}_0.
\end{align}

\begin{figure}[ht]
\centering{{\includegraphics[height=2.5 in, width=3.0 in]{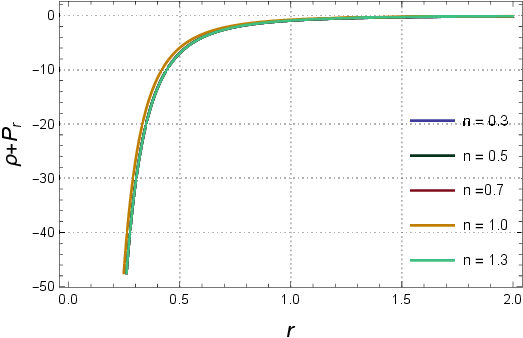}}}
\centering{{\includegraphics[height=2.5 in, width=3.0 in]{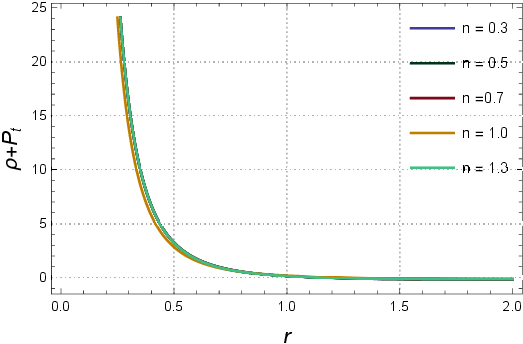}}}
\caption{NEC's behavior along radial coordinate $r$ in $f(R)$ gravity.}\label{35f}
\end{figure}

How the shape function behaves within the context of $f(R)$ gravity is depicted in Figs. \textbf{\ref{31f}} and \textbf{\ref{32f}} in Appendix B. Figure \textbf{\ref{31f}} (mentioned in Appendix B) illustrates that $b(\frac{1}{2})=\frac{1}{2}$, as shown in the left graph. The right graph in Fig. \textbf{\ref{31f}} confirms the substantiality of the well-known flaring-out limits. Moreover, Fig. \textbf{\ref{32f}} (right graph and is described in Appendix B) specifies the asymptotic flatness characteristics of the WH, while its left side graph points out that the locality of the WH throat is at $r_0 = \frac{1}{2}$, which is where the curve $b(r) - r$ intersects the $r$-axis. Moreover, Fig. \textbf{\ref{33f}} in Appendix B illustrates the variation of $b(r)$ within a particular range of Einasto index $n$ near the WH throat within the background of $f(R)$ gravity. For the shape function derived above, metric will gain the following mathematical form
\begin{align}\label{ja22}
ds^2|_{f(R)}=dt^{2}-\left\{1-\frac{1}{r({2 \alpha  \widetilde{R}+1})} \left[{\mathit{h}^3 \mathit{n} \mathcal{P}_0 \Gamma \left(3 \mathit{n},\left(\frac{\mathit{r}_0}{\mathit{h}}\right){}^{1/\mathit{n}}\right)+\mathit{r}_0+2 \alpha  \mathit{r}_0 \widetilde{R}}\right]\right\}^{-1}dr^{2}
-r^2(d\theta^{2} +\sin^2\theta{d\phi^2}).
\end{align}
These equation of states of Eq. \eqref{eoseq} provide the following relations in $f(R)$ gravity
\begin{align}\nonumber
\mathcal{W}_r|_{f(R)}&=\frac{-\exp{\left[\left(\frac{r}{\mathit{h}}\right)^{1/\mathit{n}}\right]} (\alpha  r+2 \alpha  \widetilde{R}+1)}{\mathcal{P}_0 r^3 (2 \alpha  \widetilde{R}+1)}  \left\{\mathit{h}^3 \mathit{n} \mathcal{P}_0 \left[\Gamma \left(3 \mathit{n},\left(\frac{\mathit{r}_0}{\mathit{h}}\right){}^{1/\mathit{n}}\right)-\Gamma \left(3 \mathit{n},\left(\frac{r}{\mathit{h}}\right)^{1/\mathit{n}}\right)\right]+\mathit{r}_0 (2 \alpha  \widetilde{R}+1)\right\}\\\label{ja23}&
+\frac{\alpha r}{\mathcal{P}_0 r^3 (2 \alpha  \widetilde{R}+1)},\\\nonumber
\mathcal{W}_t|_{f(R)}&= -\frac{1}{2}-\frac{1}{2r^3(2 \alpha  \widetilde{R}+1)}\left\{{\mathit{h}^3 \mathit{n} \exp{[\left(\frac{r}{\mathit{h}}\right)^{1/\mathit{n}}]} (4 \alpha  r+2 \alpha  \widetilde{R}+1) \left[\Gamma \left(3 \mathit{n},\left(\frac{r}{\mathit{h}}\right)^{1/\mathit{n}}\right)-\Gamma \left(3 \mathit{n},\left(\frac{\mathit{r}_0}{\mathit{h}}\right){}^{1/\mathit{n}}\right)\right]}\right\}\\\label{ja24}&
+\frac{\exp{[\left(\frac{r}{\mathit{h}}\right)^{1/\mathit{n}}]} \left(\mathit{r}_0 (4 \alpha  r+2 \alpha  \widetilde{R}+1)-4 \alpha  r^2\right)}{{2 r^3}\mathcal{P}_0}.
\end{align}
Figure \ref{34f} in Appendix B displays the graphs for $\mathcal{W}_{r}$ and $\mathcal{W}_{t}$ for a wide variety of galactic structures, including for the haloes of cluster size in the Millenium Run and the most massive haloes for the Millenium Simulation. As $\mathcal{W}_{r}$ approaches the WH-throat, it first rises and then decreases as it moves past this point. Figure \ref{35f} shows the regions where the NEC conditions, namely $\rho+P_r\geq0$ and $\rho+P_t\geq0$, are breached.

\section{Equilibrium Scenario and Active Gravitational Mass}

In this section, we aim to determine the conservation equation for fuzzy WH solution through the application of a shape function derived from the EDP model. The equilibrium condition will be calculated using the generalized Tolman-Oppenheimer-Volkoff (TOV) equation provided as follow \cite{albalahi2024isotropization}
\begin{align}\label{a28}
-\frac{dP_r}{dr}-\frac{a'(r)}{2}(\rho+P_r)-\frac{2}{r} (P_r-P_t)=0.
\end{align}
This equation describes the equilibrium state achieved by WH configurations, taking into account forces that can be specified under
\begin{align}\label{a29}
F_{\mathit{gf}}=-\frac{a'(r)}{2}(\rho+P_r),
\end{align}
\begin{align}\label{a30}
F_{\mathit{af}}=-\frac{2}{r} (P_r-P_t),
\end{align}
\begin{align}\label{a31}
F_{\mathit{hf}}=-\frac{dP_r}{dr}.
\end{align}
In this context, $F_{hf}$, $F_{gf}$, and $F_{af}$ represent the hydrostatic, gravitational, and anisotropic forces, respectively. The equilibrium force components-hydrostatic, gravitational and anisotropic are derived from the conservation of the effective stress–energy tensor in time independent, spherical symmetry. While the force decomposition is coordinate dependent, the physical content remains meaningful as it quantifies the balance needed to support anisotropic WH configurations in curved spacetime. The state of equilibrium is achieved when the sum of these forces becomes zero, represented as $F_{\mathit{gf}}+F_{\mathit{af}}+F_{\mathit{hf}}=0$. Specifically, in our case, the gravitational forces are negligible, leading to a simplified equilibrium condition of $F_{\mathit{a}}+F_{\mathit{h}}=0$. Through the application of $b(r)$, (i.e., shape function) resulting from the EDP model, the magnitude of the forces $F_{\mathit{a}}$ and $F_{\mathit{h}}$ have been calculated in EGB theory and are presented below
\begin{align}\nonumber
F_{\mathit{af}}&=-\frac{\left(r^2-\beta_2 \right)^2}{\alpha  r^5}+\frac{\left(\beta_2 -r^2\right) \left(r^2-\beta_2 \right)}{2 \alpha  r^5}+\frac{\beta_2 -r^2}{\alpha  r^3}
\\\label{a32}&
-\frac{\left(2 r^2-\beta_2 \right) \left(\alpha ^2+\mathit{r}_0 \left(\alpha  \mathit{h}^3+\alpha  \mathit{h}^3 \mathit{n} \rho_0 \Gamma \left(3 \mathit{n},\left(\frac{\mathit{r}_0}{\mathit{h}}\right){}^{1/\mathit{n}}\right)+r^3-\text{r$\beta_2 $}\right)+\alpha  \mathit{r}_0^2\right)}{\alpha ^3 \beta_2  r^4 \mathit{r}_0},
\end{align}
\begin{align}\label{a33}
F_{\mathit{hf}}=\frac{1}{\alpha ^3 \beta_2  r^4 \mathit{r}_0} ({3 \alpha ^5 \beta_2 -6 \alpha ^2 r^2+\mathit{r}_0 (3 \alpha ^3 \beta_2  \mathit{h}^3-6 \alpha  \mathit{h}^3 r^2
+\alpha  \mathit{h}^3 \mathit{n} \rho_0 (3 \beta_2 -6 r^2) \Gamma (3 \mathit{n},(\frac{\mathit{r}_0}{\mathit{h}}){}^{1/\mathit{n}}))
+\mathit{r}_0^2 (3 \alpha ^3 \beta_2 -6 \alpha  r^2)}),
\end{align}
where
\begin{align}\label{a34}
\beta_2=\sqrt{\alpha } \sqrt{r \left[4 \mathit{h}^3 \mathit{n} \rho_0 \Gamma \left(3 \mathit{n},\left(\frac{\mathit{r}_0}{\mathit{h}}\right){}^{1/\mathit{n}}\right)+4 \mathit{h}^3+\frac{r^3}{\alpha }+\frac{4 \alpha }{\mathit{r}_0}+4 \mathit{r}_0\right]}.
\end{align}
\begin{figure}[ht]
{{\includegraphics[height=2.5 in, width=3.0 in]{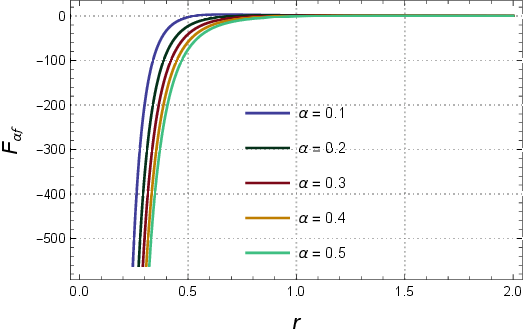}}}
{{\includegraphics[height=2.5 in, width=3.0 in]{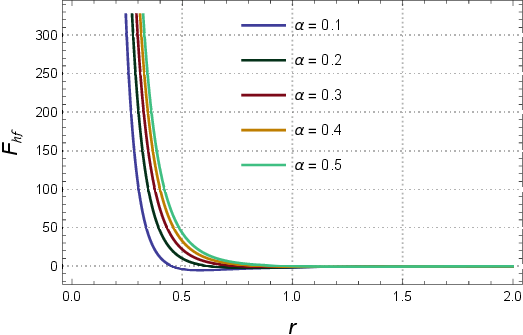}}}
{{\includegraphics[height=2.5 in, width=3.0 in]{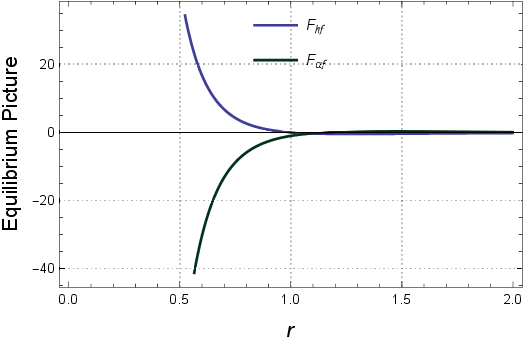}}}
\caption{The evaluation of equilibrium forces $F_{\mathit{af}}$ and $F_{\mathit{hf}}$ versus $r$ and equilibrium picture in 4D EGB theory.}\label{6f}
\end{figure}
The analysis of the aforementioned forces has been conducted in EGB theory and illustrated in Fig. \textbf{\ref{6f}}. The parameters chosen for this study include $n=0.5$, $r_0=0.5$, $h=1$, and $\rho_0=0.01$. Upon examination of Fig. \textbf{\ref{6f}}, it is noted that the forces $F_{af}$ and $F_{hf}$ exhibit similar behaviors beyond $r=2$ while acting in opposing directions. These forces appear to offset each other, indicating the potential existence of radially symmetric WH models that are not time-dependent. This inference was made based on graphs generated using specific parameter values that confirm previously discussed energy criteria in EGB theory.

The following expressions represent the magnitudes of the forces $F_{af}$ and $F_{hf}$ in $f(R)$ gravity as
\begin{align}\nonumber
  F_{\mathit{af}}|_{f(R)}&=\frac{\mathcal{P}_0 \exp[{-\left(\frac{r}{\mathit{h}}\right)^{1/\mathit{n}}}] (2 \alpha
  (r+\widetilde{R})+1)}{r^4({2 \alpha  \widetilde{R}+1})}{ \left(-r^3-3 \mathit{h}^3 \mathit{n}
  \exp[{\left(\frac{r}{\mathit{h}}\right)^{1/\mathit{n}}}] \left(\Gamma \left(3 \mathit{n},\left(\frac{r}{\mathit{h}}\right)^{1/\mathit{n}}\right)
  -\Gamma \left(3 \mathit{n},\left(\frac{\mathit{r}_0}{\mathit{h}}\right){}^{1/\mathit{n}}\right)\right)\right)} \\\label{ia29}&
 -\frac{1}{r^4}({4 \alpha  r^2+\mathit{r}_0 (6 \alpha  (r+\widetilde{R})+3)}),\\\nonumber
 F_{\mathit{hf}}|_{f(R)}&=\frac{\mathcal{P}_0 \exp[{-\left(\frac{r}{\mathit{h}}\right)^{1/\mathit{n}}}]}
 {r^4({\mathit{n}+2 \alpha  \mathit{n} \widetilde{R}})} \left(\mathit{h}^3 \mathit{n}^2 e^{\left(\frac{r}
 {\mathit{h}}\right)^{1/\mathit{n}}} (2 \alpha  r+6 \alpha  \widetilde{R}+3) \left(\Gamma \left(3 \mathit{n},\left(\frac{r}{\mathit{h}}
 \right)^{1/\mathit{n}}\right)-\Gamma \left(3 \mathit{n},\left(\frac{\mathit{r}_0}{\mathit{h}}\right){}^{1/\mathit{n}}\right)
 \right)\right)\\\label{ia30}&
+ \frac{\mathcal{P}_0 \exp[{-\left(\frac{r}{\mathit{h}}\right)^{1/\mathit{n}}}]}{r^4({\mathit{n}+2 \alpha
\mathit{n} \widetilde{R}})}\left(\alpha  r^4 \left(\left(\frac{r}{\mathit{h}}\right)^{1/\mathit{n}}-\mathit{n}\right)+\mathit{n}
 r^3 (\alpha  r+2 \alpha  \widetilde{R}+1)\right)-\frac{\mathit{r}_0}{r^4} (2 \alpha  r+6 \alpha  \widetilde{R}+3).
\end{align}
It is worth mentioning that these equations describe the fundamental forces with the present Ricci scalar choices for the wide cluster ranges of dark matter halo objects with masses ranging from dwarfs to clusters.
\begin{figure}[ht]
\centering{{\includegraphics[height=2.5 in, width=3.0 in]{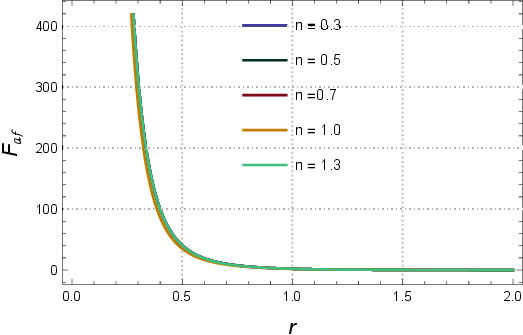}}}
\centering{{\includegraphics[height=2.5 in, width=3.0 in]{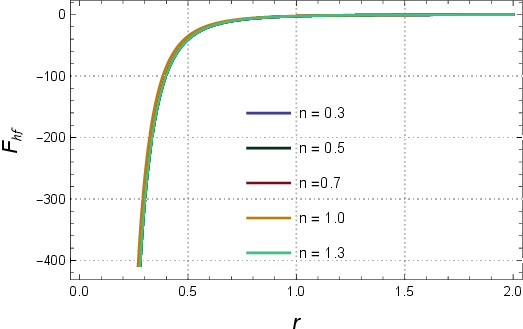}}}
\centering{{\includegraphics[height=2.5 in, width=3.0 in]{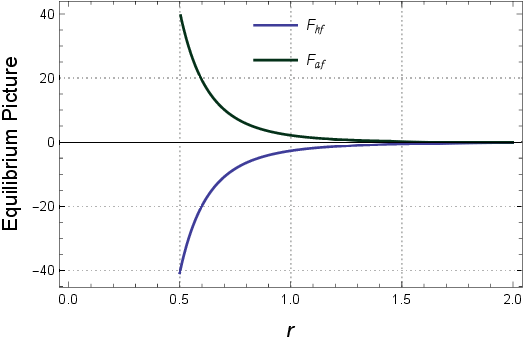}}}
\caption{The contribution of $F_{\mathit{af}}$ and $F_{\mathit{ah}}$ in the maintenance of equilibrium state of the fuzzy WH structure
in $f(R)$ gravity.}\label{36f}
\end{figure}
The forces discussed earlier have undergone analysis, and the outcomes are depicted in Fig. \textbf{\ref{36f}}. The parameters considered in the study include $r_0=0.5$, $\alpha =0.5$, $h=1$,$\widetilde{R}=0.3$ and $\mathcal{P}_0=0.1$ in $f(R)$ gravity. The examination of Fig. \textbf{\ref{36f}} indicates that despite acting in opposing directions, forces $F_{af}$ and $F_{hf}$ exhibit similar behavior. These forces appear to counterbalance each other, hinting at the feasibility of time-independent radially symmetric WH models. By utilizing graphs generated with specific parameter values that satisfy the aforementioned energy criteria, this inference was drawn. This suggests that stable fuzzy DM WH models exist in nature in the neighborhood of cold DM halos and galactic bulges in $f(R)$ gravity.

\begin{figure}[ht]
{{\includegraphics[height=2.5 in, width=3.5 in]{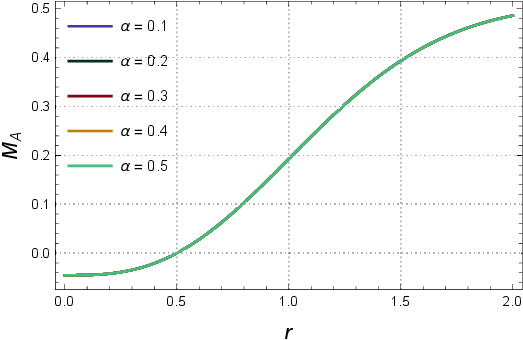}}}
\caption{The behavior of active gravitational mass of WH for $n=0.5,~h=1,~r_0=0.5$ and $\rho_0=0.01$ in 4D EGB theory.}\label{7f}
\end{figure}

We will discuss the active gravitational mass associated with fuzzy DM WHs. This mass is confined within the region of the WH from the initial throat radius $r_0$ to the boundary radius $r$. The active gravitational mass (represented by $\mathrm{M}_{\mathrm{A}}$) can be calculated using the following expression as
\begin{align}\label{a35}
\mathrm{M}_{\mathrm{A}}=4 \pi  \int_{\mathit{r}_{0}}^r \rho (r)  r^2 \, dr.
\end{align}
For fuzzy anisotropic dark matter WHs, the gravitational mass expression is found in EGB theory as under
\begin{align}\label{a36}
\mathrm{M}_{\mathrm{A}}=4 \pi  \mathit{h}^3 \mathit{n} \mathcal{\rho}_0 \left[\Gamma \left(3 \mathit{n},\left(\frac{\mathit{r}_{0}}{\mathit{h}}\right){}^{1/\mathit{n}}\right)-\Gamma \left(3 \mathit{n},\left(\frac{r}{\mathit{h}}\right)^{1/\mathit{n}}\right)\right].
\end{align}
Figure \textbf{\ref{7f}} illustrates the behavior of $\mathrm{M}_{\mathrm{A}}$ for fuzzy WH for a specific Einasto index $n=0.5$ in EGB theory.
The active gravitational mass $\mathrm{M}_{\mathrm{A}}$ diminishes with increasing $r$ due to the EDP in EGB theory. The existence of exotic matter is suggested when the active gravitational mass becomes negative in a particular spatial area, indicating a violation of energy conditions by this matter.
\begin{figure}[ht]
\centering{{\includegraphics[height=2.5 in, width=3.5 in]{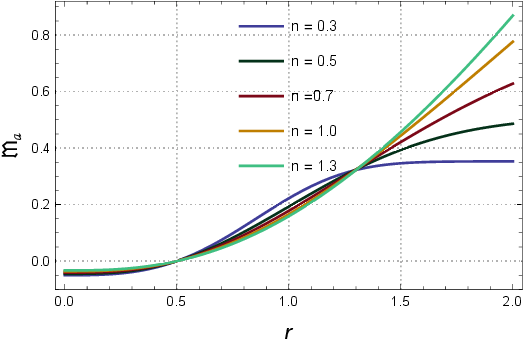}}}
\caption{The graphical analysis of active gravitational mass for various Einasto index $n$ in $f(R)$ gravity.}\label{37f}
\end{figure}

For $f(R)$ gravity setup, the gravitational mass expression is found as under
\begin{align}\label{ia32}
\mathfrak{M}_{\mathit{a}}=4 \pi  \mathcal{P}_0 \left[\mathit{n} \mathit{r}_0^3 E_{1-3 \mathit{n}}\left(\left(\frac{\mathit{r}_0}{\mathit{h}}\right){}^{1/\mathit{n}}\right)-\mathit{n} r^3 E_{1-3 \mathit{n}}\left(\left(\frac{r}{\mathit{h}}\right)^{1/\mathit{n}}\right)\right],
\end{align}
where
\begin{align}\label{ia32b}
E_{1-3 \mathit{n}}\left(\left(\frac{r}{\mathit{h}}\right)^{1/\mathit{n}}\right)
\end{align}
is the exponent integral. Here, we denote the active gravitational mass in ${f(R)}$ with $\mathfrak{M}_{\mathit{a}}$. The behavior of $\mathfrak{M}_{\mathit{a}}$ for fuzzy WH for different Einasto index $n$ as shown in Fig. \textbf{\ref{37f}}.
Because of the EDP $f(R)$ gravity, the active gravitational mass $\mathfrak{M}_{\mathit{a}}$ is negative near the wormhole throat. When the active gravitational mass in a given spatial region goes negative, it suggests the existence of exotic matter, thereby implying that this stuff is violating NEC $f(R)$ gravity. The negative active gravitational mass arises due to anisotropic pressures and the contribution of higher-curvature corrections. This effect is not unphysical in modified gravity, as it reflects deviations from the Newtonian mass definition.

\section{Complexity Factor in Fuzzy Wormhole}

\begin{figure}[ht]
\centering{{\includegraphics[height=2.5 in, width=3.0 in]{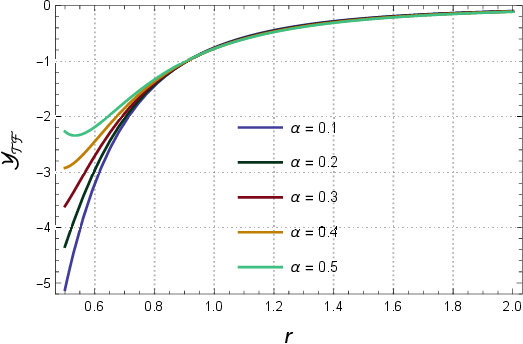}}}
\caption{The dynamics of complexity factor $\mathcal{Y}_{\mathcal{T}\mathcal{F}}$ for different values of GB parameter $\alpha$ in 4D EGB theory.}\label{8f}
\end{figure}

\begin{figure}[ht]
\centering{{\includegraphics[height=2.5 in, width=3.0 in]{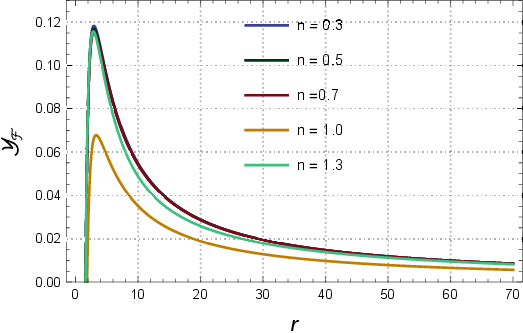}}}
\caption{The complexity factor's trajectory corresponds to different Einasto indices in $f(R)$ gravity.}\label{38f}
\end{figure}

Herrera proposed the concept of a complexity factor for time-independent, spherically symmetric compact objects in 2018 \cite{herrera2018new}. The notion of the complexity factor is predicated on systems that are simple or minimally complex and have uniform energy density and pressure that are isotropic. This sort of fluid distribution has a non-existent (or zero) complexity factor. Bondi, axial metrics, and other cosmic stellar solutions are studied using the complexity factor for this type of matter distribution \cite{herrera2019complexity,herrera2020stability,contreras2022uncharged,herrera2024cracking,herrera2024irreversibility}. Furthermore, the complexity factor of systems comprising compact objects exhibiting inhomogeneous energy density and anisotropic pressure diminishes to zero as the individual impacts of these characteristics nullify each other \cite{maurya2022relativistic,yousaf2020definition,andrade2022stellar,leon2023spherically,andrade2023anisotropic,mazharimousavi2009effect}. The definition of the complexity factor is provided as follows \cite{herrera2018new}
\begin{align}\label{a37}
\mathcal{Y}_{\mathcal{T}\mathcal{F}} = ({P_r}-{P_t})-\frac{1}{2 r^3} {\int_{\mathit{r}_{0}}^r r^3  \rho'(r)  \, dr}.
\end{align}
We have obtained a complexity factor of our system from the above formula in both EGB and $f(R)$ theories. After some manipulation, the final versions of the complexity factor are mentioned in Appendix A. The complexity trend of fuzzy DM WH along radial axis $r$ for diverse values of GB parameter is depicted in Fig. \textbf{\ref{8f}} for EGB theory. In the context of $f(R)$ theory, the behavior is described in Fig. \textbf{\ref{38f}}. It has been observed that as $r$ moves towards infinity or away from the WH throat, $\mathcal{Y}_{\mathcal{T}\mathcal{F}}$ tends towards zero. The lowest complexity factor signifies isotropic pressure and homogeneous energy density, which justifies our results in comparison to \cite{herrera2018new}. Furthermore, non-uniform energy distribution and directional pressure can both be predicted by the zero complexity element until the opposing effects of these two elements achieve equilibrium in terms of complexity in EGB theory. Consequently, as one increases radial coordinates, $\mathcal{Y}_{\mathcal{T}\mathcal{F}}$ drops towards zero, and the complexity element gradually grows towards the WH throat. In addition, the complexity dynamics place more importance on pressure uniformity than on energy density homogeneity. Although a reduction in complexity suggests simpler equilibrium structures, a conclusive assessment of stability requires dynamical perturbation analysis. We propose this as a future extension, particularly through linearized perturbations around the static background.

\section{Conclusions}

It is widely recognized that for a traditional WH model to be feasible, a crucial element must exist that violates the energy condition (EC) within the model's constraints. When examining modified gravitational models as a theoretical framework, the effective stress-energy tensor within WH passages adheres to ECs, potentially enabling the creation of distinct WH structures with additional curvature modifications. This research explores the spherically symmetric fuzzy DM WH solutions linked to anisotropic matter distribution within the frameworks of 4D EGB and $f(R)$ gravity models. Our involvement in EGB and $f(R)$ models stems from their significant coverage in the scientific literature concerning cosmic phenomena such as black holes, wormholes, and stellar formations. We have explored the potential for modeling central galactic structures using EGB and $f(R)$ models of gravity by proposing fuzzy WHs composed of DM. Our analysis has examined how the NEC impacts the distribution of matter content, particularly anisotropic fluids, to assess the effect of the EDP on the existence of WH models. In the existing literature, several established methods are available for analyzing WH models. One method involves hypothesizing the shape function for WH and studying ECs, while another method focuses on determining the shape function by assuming fluid components with strong coherence. By aligning the field equation with Einasto dark matter energy density and integrating the derived equation, the shape function is obtained. The evolution of the shape function is further examined and visually depicted through graphical representations (\textbf{\ref{1f}}, \textbf{\ref{2f}}, \textbf{\ref{3f}}, \textbf{\ref{31f}}, \textbf{\ref{32f}} and \textbf{\ref{33f}}, mentioned in Appendix B). Various plots have been created to highlight numerical outcomes related to the shape function, showcasing the adherence to throat and flare-out conditions, which are also visually illustrated. Through these figures, we conclude that the shape
function derived from EDP could effectively define the geometry of the WH, as it satisfies the flare-out conditions,
throat conditions, and flatness criteria in both EDP and $f(R)$ gravity.

The development of tangential and radial EOS has also been examined for the different choices of EGB. When an observer resides away from the WH throat, the value of $\mathcal{W}_{\mathit{r}}$ decreases adversely, yet close to $r_0$, we observe growing behavior. In contrast, a different pattern is observed with $\mathcal{W}_{\mathit{t}}$ as described by Fig. \textbf{\ref{4f}} in Appendix B.  The evolution of tangential and radial EOS has also been explored for a pattern of halo objects including cold dark matter halos, spiral galactic bulges, and cluster-sized haloes in the Millenium Run depending upon the choice of index $n$ in $f(R)$ gravity. Away from the WH throat, the value of $\mathcal{W}_{1}$ decreases unfavorably, yet close to $r_0$, an increasing pattern is observed.
Conversely, a different pattern is observed with $\mathcal{W}_{2}$ as shown in Fig. \textbf{\ref{34f}} in Appendix B.

Numerous methods exist in the literature for studying WH configurations, including defining the
shape function and analyzing the NEC. Our study focuses on NEC's behavior, revealing dissatisfaction at the WH throat $(r_0 = 0.5)$ based on valid areas and NEC plots. An examination of the active gravitational mass indicates a negative value, implying the existence of exotic matter that contradicts ECs. Additionally, an analysis of the equilibrium forces for fuzzy WHs includes the computation of hydrostatic and anisotropic forces. The plots \textbf{\ref{5f}} and \textbf{\ref{35f}} demonstrate how they entirely cancel each other out. It is concluded that within a proper arena of parametric Einasto choices, the WEC is dissatisfied thus indicating the occurrences of exotic matter for fuzzy dark matter WHs both in EGB and $f(R)$
gravity (indicated in Figs. \textbf{\ref{5f}} and \textbf{\ref{35f}}).

We have performed an analysis of the equilibrium forces for fuzzy WHs by evaluating
hydrostatic and anisotropic forces. The stable and well-sustained geometries of fuzzy dark
matter WHs are available in nature under the effects of extra curvature $f(R)$ terms as
described by Figs. \textbf{\ref{6f}} and \textbf{\ref{36f}}. The contribution of active matter content quantity in producing gravitational
interaction in EGB and $f(R)$ gravity keeps on increasing with the radial coordinate $r$ in an atmosphere of a variety of astrophysical systems mediated by the Einasto index $n$. This can be well justified from Figs. \textbf{\ref{7f}} and \textbf{\ref{37f}}.

The FDM-supported WHs may produce gravitational lensing features deviating from both Schwarzschild and soliton-core profiles. The observable signatures include central brightness suppression, deflection angles deviating from Schwarzschild behavior, and modified photon spheres \cite{rahaman2014possible,event2019first}. We estimate the throat radius to be $\sim 0.1-1 kpc$, within dense galactic centers, consistent with soliton cores of FDM models. The FDM WHs may exhibit astrophysical signatures distinct from both solitonic cores and classical black holes. Deviations in deflection angle, photon sphere size, or shadow shape may arise due to the non-singular, anisotropic nature of the WH geometry. These could mimic or distort the shadow of a black hole (e.g., EHT observations) \cite{tsukamoto2016strong,bambi2013broad}. The light traversing different paths through a wormhole could produce exotic time-delay structures or duplicated images in strong lensing systems. The WH's throat may introduce local anomalies in inner-galactic rotation curves, particularly for low Einasto-index cores. Future high-resolution imaging (e.g., ngEHT, SKA), pulsar timing arrays, or microlensing surveys may help constrain such possibilities \cite{rahaman2014possible}. We have taken into consideration the static irrotational spacetime metric of the WH and subsequently deduced the complexity factor, motivated by the work by Herrera \cite{herrera2018new}.
The complexity factor of fuzzy dark matter WHs is calculated, displaying a consistent increase over time. On increasing the contribution of EGB theory controlled by $\alpha$, the value of the complexity factor increases. Thus EGB theory is likely to host more complex systems. However, in the context of $f(R)$ gravity indicates that the complexity takes on its role in the structure of WH, and after attaining a specific radial coordinate value (i.e., $r=4.9$), the contribution of complexity reduces gradually.
This clearly shows that less-complex WH geometric within the framework of specific halo objects exist, as an observer moves away from the central core of the spherical geometric objects in $f(R)$ gravity. The important physics understood from our analysis is that the realistic geometries of fuzzy dark matter WHs exist in nature in the surroundings of various galactic haloes in both EGB and $f(R)$ gravity. It would be worthwhile to analyze the complexity-free analysis for the diffused fuzzy black hole droplets by using our presented technique which could be our future work. Our analysis focuses on FDM; other interacting DM models, such as mirror DM \cite{oikonomou2024low}, also lead to modified stress-energy tensors and anisotropic pressures, both of which are key ingredients in constructing traversable wormholes in modified gravity. Future work may extend our methods to include such alternative dark sector models.

\section*{Acknowledgement}

The work of KB was supported by the JSPS KAKENHI Grant Numbers 21K03547, 23KF0008, 24KF0100 and Competitive Research Funds for Fukushima University Faculty (25RK011). The work by BA was supported by the Ongoing Research Funding program (ORF-2025-464), King Saud University, Riyadh, Saudi Arabia.

\section*{Conflict of Interest}

The authors declare no conflict of interest.

\section*{Data Availability Statement}

This manuscript has no associated data
or the data will not be deposited. [Authors comment: This manuscript
contains no associated data.]
\renewcommand{\theequation}{A\arabic{equation}}
\setcounter{equation}{0}
\section*{Appendix A: Expressions of Complexity Factor in both Gravity Models}

In this Appendix, we provide the expressions of complexity factor obtained in the context of gravity models under consideration. The following complexity factor has been achieved by using the values of our under-considered system in Eq. \eqref{a37}. This is found for 4D EGB theory as
\begin{align}\nonumber
\mathcal{Y}_{\mathcal{T}\mathcal{F}} &= -\frac{1}{2 r^3} \Bigg\{-\frac{2}{\alpha} \Bigg(r^3 + \frac{\alpha^2}{\mathit{r}_0} + \alpha \mathit{r}_0 - \beta_1 r
+ \alpha \mathit{h}^3 \mathit{n} \rho_0
\Bigg[\Gamma\Big(3 \mathit{n}, \Big(\frac{\mathit{r}_0}{\mathit{h}}\Big)^{1/\mathit{n}}\Big)
- \Gamma\Big(3 \mathit{n}, \Big(\frac{r}{\mathit{h}}\Big)^{1/\mathit{n}}\Big)\Bigg]\Bigg)
\\\nonumber
&\quad + \frac{\Big(2 r^2 - 3 \beta_1 \Big)
\Bigg(\alpha^2 + \alpha \rho_0 \mathit{h}^3 \mathit{n} \mathit{r}_0
\Bigg[\Gamma\Big(3 \mathit{n}, \Big(\frac{\mathit{r}_0}{\mathit{h}}\Big)^{1/\mathit{n}}\Big)
- \Gamma\Big(3 \mathit{n}, \Big(\frac{r}{\mathit{h}}\Big)^{1/\mathit{n}}\Big)\Bigg]
+ 2 \mathit{r}_0^2\Bigg)}{\alpha^3 \beta_1 \mathit{r}_0}
\\\nonumber
&\quad + \frac{\alpha \rho_0 \mathit{r}_0 r^3 \Big(\beta_1 - 2 r^2\Big)
e^{-\Big(\frac{r}{\mathit{h}}\Big)^{1/\mathit{n}}}
- r^3 \mathit{r}_0 \Big(r^2 - \beta_1 \Big)}{\alpha^3 \beta_1 \mathit{r}_0}
\\\label{a38}
&\quad + \mathit{h}^3 \rho_0
\Bigg[\Gamma\Big(3 \mathit{n} + 1, \Big(\frac{r}{\mathit{h}}\Big)^{1/\mathit{n}}\Big)
- \Gamma\Big(3 \mathit{n} + 1, \Big(\frac{\mathit{r}_0}{\mathit{h}}\Big)^{1/\mathit{n}}\Big)\Bigg]\Bigg\},
\end{align}
while for $f(R)$ gravity, it is found to be
\begin{align}\nonumber
\mathcal{Y}_{\mathcal{F}}|_{f(R)} &= \frac{1}{2 r^3} \left\{
\mathcal{P}_0 \mathit{r}_0^3 \left(\frac{\mathit{r}_0}{\mathit{h}}\right)^{1/\mathit{n}}
E_{-3 \mathit{n}}\left(\left(\frac{\mathit{r}_0}{\mathit{h}}\right)^{1/\mathit{n}}\right)
+ \mathcal{P}_0 \left(-r^3\right) \left(\frac{r}{\mathit{h}}\right)^{1/\mathit{n}}
E_{-3 \mathit{n}}\left(\left(\frac{r}{\mathit{h}}\right)^{1/\mathit{n}}\right)
\right\} \\ \nonumber
&\quad + \frac{\mathcal{P}_0 \exp\left\{\left[-\left(\frac{r}{\mathit{h}}\right)^{1/\mathit{n}}\right]\right\}
\left[2 \alpha r + 2 \alpha R + 1\right]}{2 r^3 \left(2 \alpha R + 1\right)}
\left\{ r^3 - 3 \mathit{h}^3 \mathit{n} \exp\left\{\left[\left(\frac{r}{\mathit{h}}\right)^{1/\mathit{n}}\right]\right\}
\Gamma\left(3 \mathit{n}, \left(\frac{\mathit{r}_0}{\mathit{h}}\right)^{1/\mathit{n}}\right)
\right\} \\ \nonumber
&\quad + \frac{\mathcal{P}_0 \exp\left\{\left[-\left(\frac{r}{\mathit{h}}\right)^{1/\mathit{n}}\right]\right\}
\left[2 \alpha r + 2 \alpha R + 1\right]}{2 r^3 \left(2 \alpha R + 1\right)}
\left\{ 3 \mathit{h}^3 \mathit{n}
\Gamma\left(3 \mathit{n}, \left(\frac{r}{\mathit{h}}\right)^{1/\mathit{n}}\right) \right\} \\ \label{ia34}
&\quad + \frac{1}{2 r^3} \left\{ 4 \alpha r^2 - 3 \mathit{r}_0
\left(2 \alpha r + 2 \alpha R + 1\right)\right\}.
\end{align}
We evaluate a constraint for obtaining a less complex WH structure by assigning a zero value to the above equation in our analysis.
\renewcommand{\theequation}{B\arabic{equation}}
\setcounter{equation}{0}
\section*{Appendix B: Analysis of Shape Functions in both Gravity Models}
\begin{figure}[ht]
\centering{{\includegraphics[height=2.5 in, width=3.0 in]{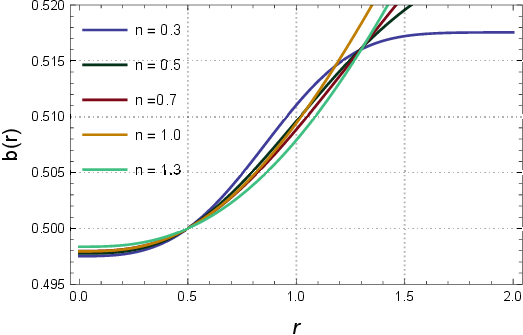}}}
\centering{{\includegraphics[height=2.5 in, width=3.0 in]{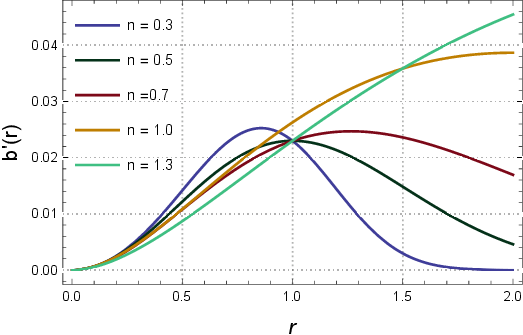}}}
\caption{The trajectory of $b(r)$ and $b'(r)$ versus $r$ axis in $f(R)$ gravity.}\label{31f}
\end{figure}

\begin{figure}[ht]
\centering{{\includegraphics[height=2.5 in, width=3.0 in]{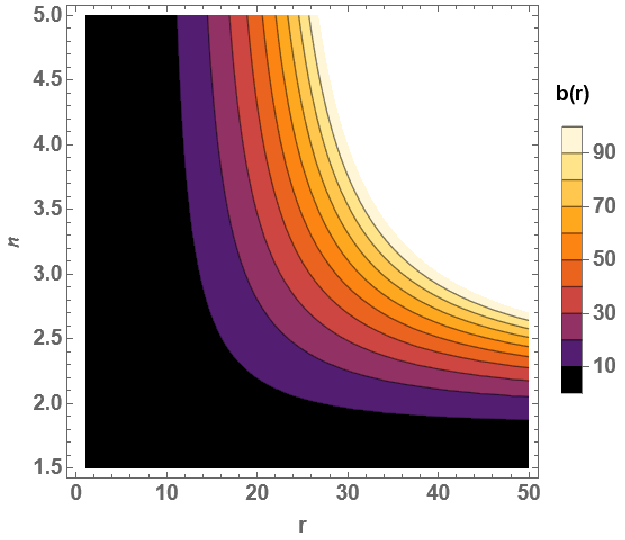}}}
\centering{{\includegraphics[height=2.5 in, width=3.0 in]{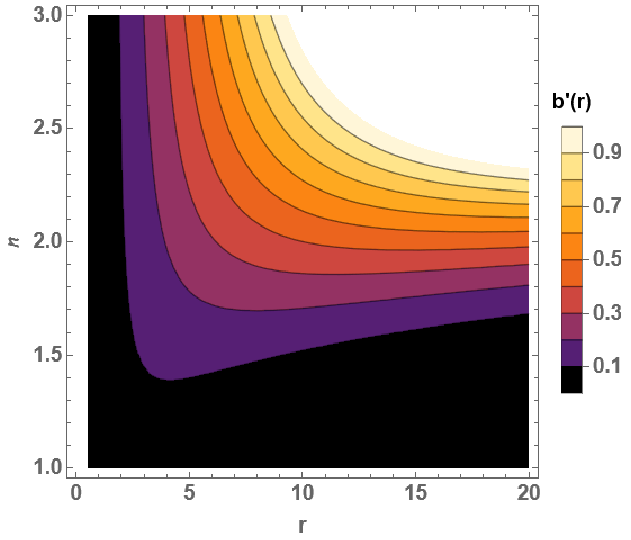}}}
\caption{The contour plots of $b(r)$ and $b'(r)$ for certain domain of $n$ in $f(R)$ gravity.}\label{33f}
\end{figure}

\begin{figure}[ht]
\centering{{\includegraphics[height=2.5 in, width=3.0 in]{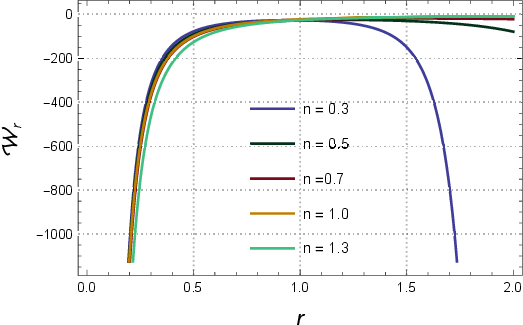}}}
\centering{{\includegraphics[height=2.5 in, width=3.0 in]{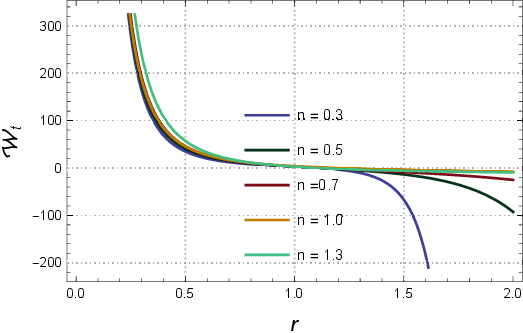}}}
\centering{{\includegraphics[height=2.5 in, width=3.0 in]{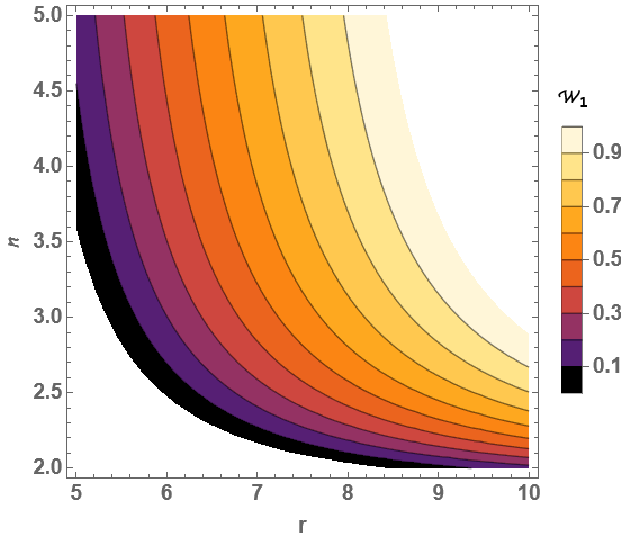}}}
\centering{{\includegraphics[height=2.5 in, width=3.0 in]{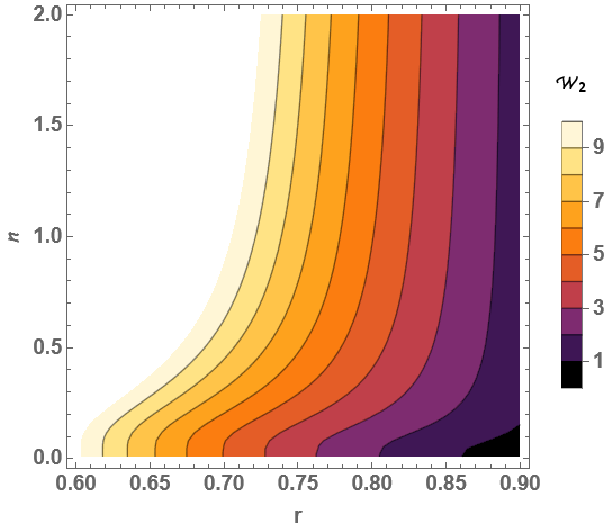}}}
\caption{$\mathcal{W}_{r}$ and $\mathcal{W}_{t}$ along radial axis $r$ for some values of Einasto index $n$ in $f(R)$ gravity.}\label{34f}
\end{figure}

In this Appendix, we study the behavior of the shape functions obtained in the context of gravity models under consideration, for different values of parameters involved.
\begin{figure}[ht]
{{\includegraphics[height=2.5 in, width=3.0 in]{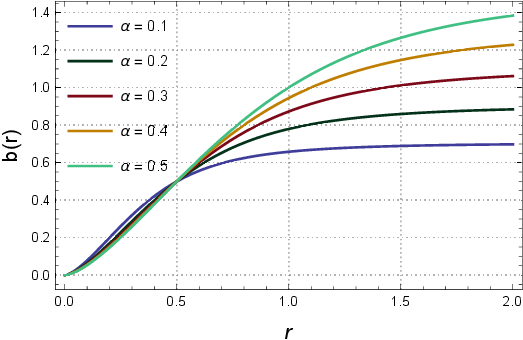}}}
{{\includegraphics[height=2.5 in, width=3.0 in]{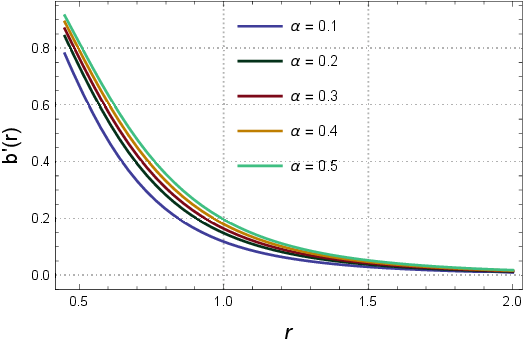}}}
\caption{The relation between $b(r)$ and $r$ in 4D EGB theory is shown in the left figure. The picture on the right side displays the trajectory of $b'(r)<1$. In this instance, $r_0=0.5$.}\label{1f}
\end{figure}
\begin{figure}[ht]
{{\includegraphics[height=2.5 in, width=3.0 in]{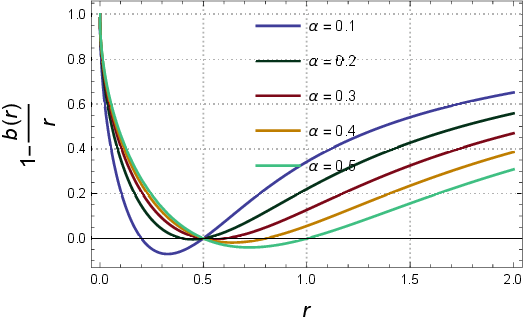}}}
{{\includegraphics[height=2.5 in, width=3.0 in]{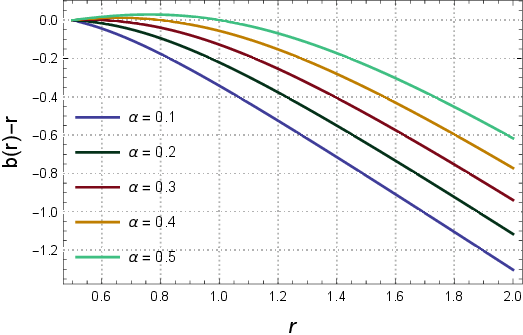}}}
\caption{The dynamics of $1-\frac{b(r)}{r}$ and $b(r)-r$ versus $r$ in 4D EGB theory. }\label{2f}
\end{figure}
\begin{figure}[ht]
{{\includegraphics[height=2.5 in, width=3.0 in]{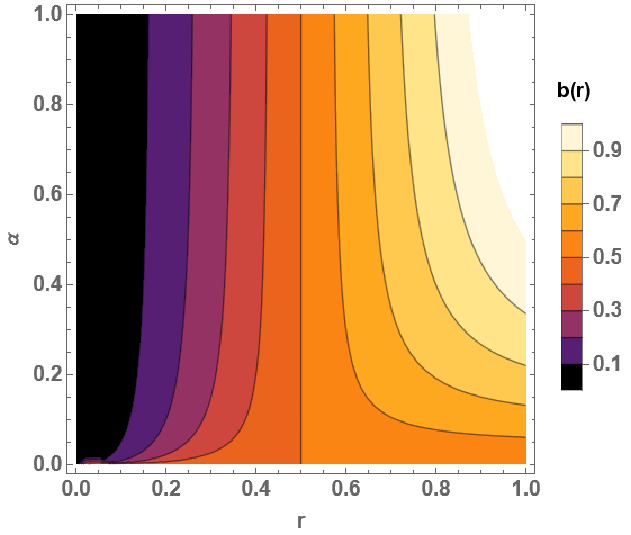}}}
{{\includegraphics[height=2.5 in, width=3.0 in]{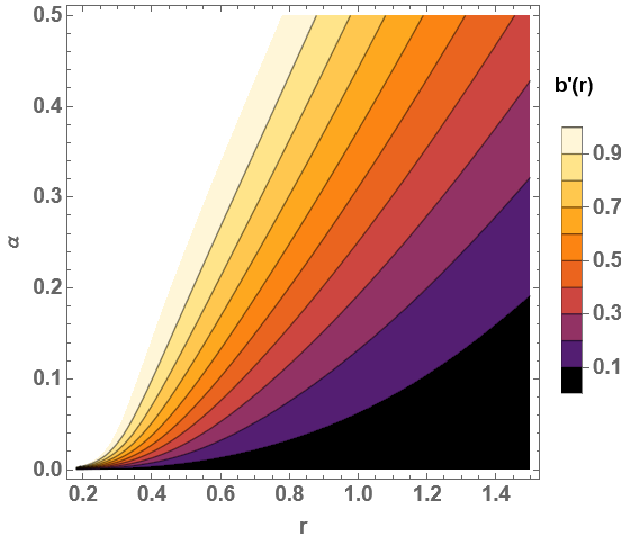}}}
\caption{The contour plots of $b(r)$ and $b'(r)$ corresponding to a particular $\alpha$ domain in 4D EGB theory.}\label{3f}
\end{figure}
 \begin{figure}[ht]
{{\includegraphics[height=2.5 in, width=3.0 in]{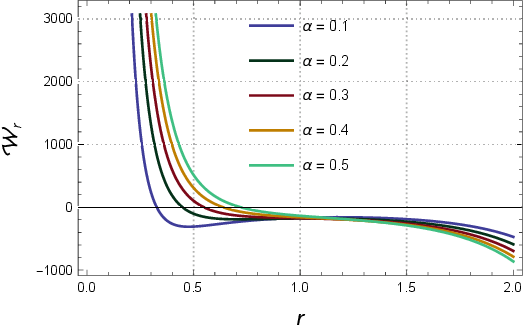}}}
{{\includegraphics[height=2.5 in, width=3.0 in]{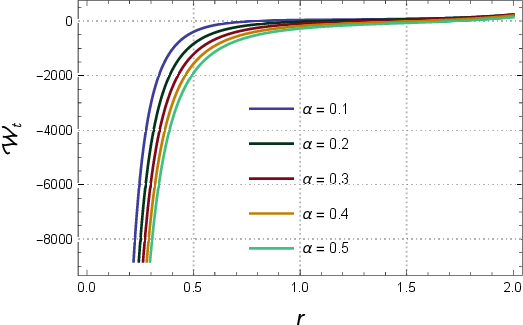}}}
{{\includegraphics[height=2.5 in, width=3.0 in]{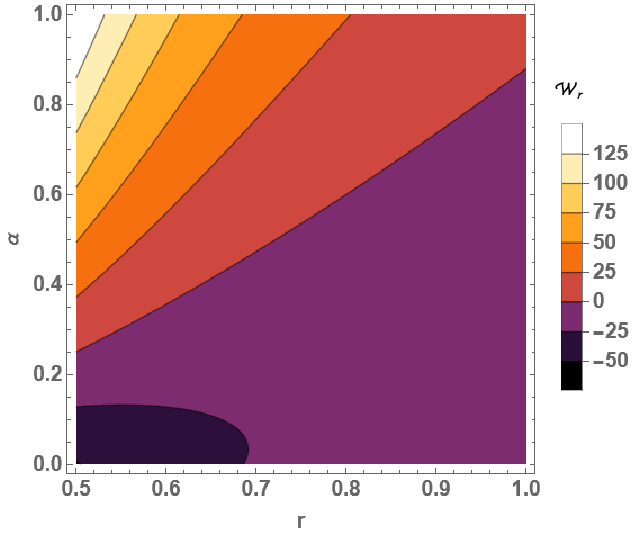}}}
{{\includegraphics[height=2.5 in, width=3.0 in]{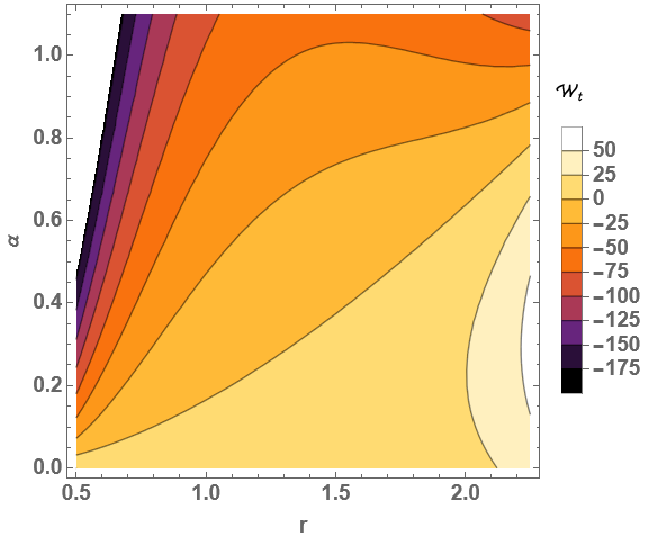}}}
\caption{The behavior of $\mathcal{W}_{\mathit{r}}$ and $\mathcal{W}_{\mathit{t}}$ along radial axis $r$ for different choices of $\alpha$ in 4D EGB theory.}\label{4f}
\end{figure}The left diagrams of Figs. \textbf{\ref{1f}}, \textbf{\ref{31f}}, \textbf{\ref{3f}} and \textbf{\ref{33f}} describe the diagrams of the shape function of the WH. One can notice elaborately the properties related to the structure of WH via these graphs. The location of the WH throat, where the radial coordinate achieves the least value, is shown by the graph's smallest value of $b(r)$. This also describes the transition of the WH from its neck to the surrounding space away from the throat (as $r$ decreases or increases) in the subsequent theories. The minimum value of $b(r)$ in the same graph points to the throat size thereby providing more insights insight into the size and stability of the traversable WH epochs. The variations in the spherically symmetric WH radius (or its effective size) in both EGB and $f(R)$ theories versus if an observer moves along the radial direction can be visualized through the left graphs of Figs. \textbf{\ref{1f}}, \textbf{\ref{31f}}, \textbf{\ref{3f}} and \textbf{\ref{33f}}. The rate of radial change in the shape function versus the radial coordinate is shown in the right diagrams of Figs. \textbf{\ref{1f}}, \textbf{\ref{31f}}, \textbf{\ref{3f}} and \textbf{\ref{33f}}. No spikes and sudden variations in $b'(r)$ at certain values of $r$ after including corrections from EGB and $f(R)$ theories are observed, thereby providing feasible and realistic WH features including throat's least radius or transitions in the geometry of WH. A smooth and continuous trajectories of $b'(r)$ are noticed from the right diagrams of Figs. \textbf{\ref{1f}}, \textbf{\ref{31f}}, \textbf{\ref{3f}} and \textbf{\ref{33f}} thereby pointing out relatively stable WH with the potentially traversable geometry. The negative regions of $b(r)-r$ have been shown for certain radial coordinate values in Figs. \textbf{\ref{3f}} and \textbf{\ref{33f}}. These graphs indicate that the observer observes a collapsing/pinching off of the spatial spherically symmetric dimensions when an observer moves away from WH throat, i.e., $(r>r_0)$. This further leads to an understanding that our fuzzy dark matter WH is not traversable, thus reinforcing that some sort of exotic matter is required to stabilize WH geometry. This reveals the requirement of the exotic matter for its stability that would prevent our WH structure from collapsing. Figures \textbf{\ref{4f}} and \textbf{\ref{34f}} describe that $\mathcal{W}_{\mathit{r}}$ and $\mathcal{W}_{\mathit{t}}$ are within the feasible bounds.
\begin{figure}[ht]
\centering{{\includegraphics[height=2.5 in, width=3.0 in]{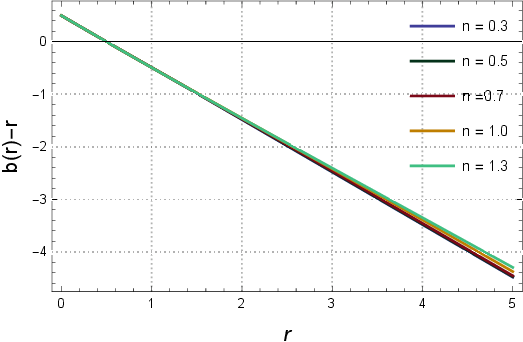}}}
\centering{{\includegraphics[height=2.5 in, width=3.0 in]{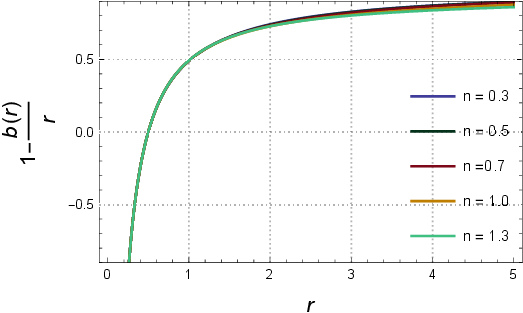}}}
\caption{The trajectory of $b(r)-r$ and $1-\frac{b(r)}{r}$ along $r$ in $f(R)$ gravity.}\label{32f}
\end{figure}
The behavior of the shape function is depicted in Figs. \textbf{\ref{31f}} and \textbf{\ref{32f}}. Figure \textbf{\ref{31f}} illustrates that $b(\frac{1}{2})=\frac{1}{2}$, as shown in the left graph. The right graph in Fig. \textbf{\ref{31f}} confirms the substantiality of the well-known flaring-out limits. Moreover, Fig. \textbf{\ref{32f}} (right graph) specifies the asymptotic flatness characteristics of the WH, while its left side graph points out that the locality of the WH throat is at $r_0 = \frac{1}{2}$, which is where the curve $b(r) - r$ intersects the radial axis. Moreover, Fig. \textbf{\ref{33f}} illustrates the variation of $b(r)$ within a particular range of Einasto index $n$ near the WH throat.

\end{document}